\documentclass[aip, jap, amsmath, amssymb,reprint]{revtex4-1}

\usepackage{graphicx}
\usepackage{dcolumn}
\usepackage{bm}

\usepackage[utf8]{inputenc}
\usepackage[T1]{fontenc}
\usepackage{mathptmx}
\usepackage{etoolbox}

\makeatletter
\def\@email#1#2{%
 \endgroup
 \patchcmd{\titleblock@produce}
  {\frontmatter@RRAPformat}
  {\frontmatter@RRAPformat{\produce@RRAP{*#1\href{mailto:#2}{#2}}}\frontmatter@RRAPformat}
  {}{}
}%
\makeatother

\begin{document}

\title[Thermal relaxation of strain and twist in ferroelectric hexagonal boron nitride moir{\'e} interfaces]{Thermal relaxation of strain and twist\\in ferroelectric hexagonal boron nitride moir{\'e} interfaces}

\author{Marisa Hocking}
\affiliation{Department of Materials Science and Engineering, Stanford University, Stanford, CA 94305}
\affiliation{Stanford Institute for Materials and Energy Sciences, SLAC National Accelerator Laboratory, Menlo Park, CA 94025}

\author{Christina E. Henzinger}
\affiliation{Department of Materials Science and Engineering, Stanford University, Stanford, CA 94305}

\author{Steven J. Tran}
\affiliation{Stanford Institute for Materials and Energy Sciences, SLAC National Accelerator Laboratory, Menlo Park, CA 94025}
\affiliation{Department of Physics, Stanford University, Stanford, CA 94305}

\author{Mihir Pendharkar}
\affiliation{Department of Materials Science and Engineering, Stanford University, Stanford, CA 94305}
\affiliation{Stanford Institute for Materials and Energy Sciences, SLAC National Accelerator Laboratory, Menlo Park, CA 94025}

\author{Nathan J. Bittner}
\affiliation{Independent Researcher}

\author{Kenji Watanabe}
\affiliation{Research Center for Electronic and Optical Materials, National Institute for Materials Science, 1-1 Namiki, Tsukuba 305-0044, Japan}

\author{Takashi Taniguchi}
\affiliation{Research Center for Materials Nanoarchitectonics, National Institute for Materials Science,  1-1 Namiki, Tsukuba 305-0044, Japan}

\author{David Goldhaber-Gordon}
\affiliation{Stanford Institute for Materials and Energy Sciences, SLAC National Accelerator Laboratory, Menlo Park, CA 94025}
\affiliation{Department of Physics, Stanford University, Stanford, CA 94305}

\author{Andrew J. Mannix}
\email{ajmannix [at] stanford [dot] edu}
\affiliation{Department of Materials Science and Engineering, Stanford University, Stanford, CA 94305}
\affiliation{Stanford Institute for Materials and Energy Sciences, SLAC National Accelerator Laboratory, Menlo Park, CA 94025}

\date{\today}

\begin{abstract}
New properties can arise at van der Waals (vdW) interfaces hosting a moir{\'e} pattern generated by interlayer twist and strain. However, achieving precise control of interlayer twist/strain remains an ongoing challenge in vdW heterostructure assembly, and even subtle variation in these structural parameters can create significant changes in the moir{\'e} period and emergent properties. Characterizing the rate of interlayer twist/strain relaxation during thermal annealing is critical to establish a thermal budget for vdW heterostructure construction and may provide a route to improve the homogeneity of the interface or to control its final state. Here, we characterize the spatial and temporal dependence of interfacial twist and strain relaxation in marginally-twisted hBN/hBN interfaces heated under conditions relevant to vdW heterostructure assembly and typical sample annealing. We find that the ferroelectric hBN/hBN moir{\'e} relaxes minimally during annealing in air at typical assembly temperatures of 170ºC. However, at 400°C, twist angle relaxes significantly, accompanied by a decrease in spatial uniformity. Uniaxial heterostrain initially increases and then decreases over time, becoming increasingly non-uniform in direction. Structural irregularities such as step edges, contamination bubbles, or contact with the underlying substrate result in local inhomogeneity in the rate of relaxation.

\end{abstract}

\maketitle

\section{\label{sec1:level1}Introduction}

The weak van der Waals bonding between layers of 2D materials such as graphene and hexagonal boron nitride (hBN) facilitates the assembly of heterostructures. For interfaces between two layers, a twist angle or small difference in lattice constant will result in a moir{\'e} superlattice. This decoupling from the atomic lattice permits the fabrication of unprecedented structures which can exhibit phenomena including unconventional superconductivity,\cite{Cao2018-sl} orbital ferromagnetism,\cite{Sharpe2019-ot} and Wigner states.\cite{Regan2020-pd}

In monolayers which lack inversion symmetry, such as WSe\textsubscript{2},\cite{Ko2023-zt, Deb2022-zw} MoS\textsubscript{2},\cite{Deb2022-zw, Weston2022-ju} and hBN,\cite{Woods2021-rz, Yasuda2021-cs, Vizner_Stern2021-gw} out-of-plane ferroelectricity emerges for stacking sequences which break inversion symmetry. For perfect rotational alignment between layers, this generates a single ferroelectric domain. In practice, targeting zero twist when assembling a vdW stack produces a finite near-zero twist angle – the “marginally-twisted” regime – resulting in a moir{\'e} lattice which relaxes into ferroelectric domains (two per moir{\'e} unit cell, with one polarity upward and one polarity downward).\cite{Weston2022-ju, Vizner_Stern2021-gw, Yasuda2021-cs} The domain boundaries – variously referred to as soliton domain boundaries or partial dislocations – will form a perfect dislocation upon meeting and can only be eliminated by reaching an edge of the sample.\cite{Ko2023-zt}  Applying an out-of-plane electric field causes the ferroelectric domains with polarity aligned along the applied field to expand and anti-aligned domains to contract via bowing of the domain walls, but complete nucleation or annihilation of an individual domain is difficult.\cite{Weston2022-ju, Ko2023-zt} 

Previous works have used the spatial modulation of electrostatic potential in ferroelectric hBN to modify the opto-electronic properties of an adjacent 2D semiconductor (i.e., a “proximity moir{\'e}” effect), revealing distinct excitonic species in regions with opposite polarization\cite{Fraunie2023-gj} as well as confining exciton diffusion lengths.\cite{Kim2023-lg} However, characterizing the optical excitations associated with a single domain requires individual domains that are optically addressable, i.e., at least several hundred nanometers wide. This necessitates achieving twist angles within a few tenths of a degree from zero (e.g., < 0.03° for domains to reach > 500 nm). Current heterostructure assembly techniques cannot reliably reach this small angle, typically deviating from the intended angle due to twist relaxation and unintentional mechanical forces during assembly. Post-assembly modification of moir{\'e} superlattices is an appealing alternative route to increase yield of samples with large domains of alternating polarizations.

The interface between two incommensurate flakes may display superlubricity – easy slippage with respect to each other.\cite{Yang2020-gk} Thus, applying mechanical force after stacking has succeeded in producing macroscopic rotation.\cite{Yang2020-gk, Du2017-eg, Wang2016-yj, Ribeiro-Palau2018-mu, Kapfer2023-ji} However, interfaces nearing commensurability (as at the low twist angles we’ve described targeting) may be locked into place,\cite{Yang2020-gk} preventing further rotation without damaging the material. Another limitation is that applying forces to drive rotation has typically involved contacting the sample with an external probe, which may increase contamination on the surface or deform the flakes.

High-temperature annealing after assembly has been shown to alter interfacial twist.\cite{Wang2016-yj, Souibgui2017-ma, Baek2023-ic} Previous experimental efforts have shown there is a maximum critical angle ($\theta_c$ > 1°) below which the interface relaxes towards 0°\cite{Wang2016-yj, Baek2023-ic} as well as a critical temperature (T$_c$ $\geq$ 600°C) at which incommensurate regions begin moving or are entirely driven out.\cite{Baek2023-ic, Zhu2018-ya, Alden2013-mi, Woods2016-lb, Wang2015-ly} Relaxation below T$_c$ has been less frequently studied with results ranging from a linear reduction in twist angle\cite{Wang2016-yj} to local fluctuation of an incommensurate domain wall without macroscopic rotational relaxation.\cite{De_Jong2022-fo}  Reports suggest that the interfacial twist becomes more (less) uniform for heterostructures with large (small) initial angles after annealing below T$_c$,\cite{Woods2016-lb, Zhu2016-mq} though this is commonly extracted from fingerprint Raman peaks, which is an optical technique with a spot size on the order of several hundred nanometers. Moreover, the effect of annealing on strain, which heavily influences the geometry of the moir{\'e} as well as electronic behavior, is not well understood. 

Here, we study the effects of thermal annealing upon marginally twisted ferroelectric hBN/hBN interfaces. We create a series of twisted hBN samples with angles $\leq$ 0.3°. They are annealed at elevated temperatures in both air and vacuum for sustained periods of time. Before and after annealing, the real-space moir{\'e} is mapped using scanning probe techniques. We extract the nominal interlayer twist and uniaxial heterostrain within each region by applying real-space and reciprocal-scape analysis techniques. At 400°C, we find a reduction in twist that saturates at longer annealing times. Uniaxial heterostrain typically increases but then reduces again. These changes are spatially inhomogeneous and are sensitive to structural irregularities such as step edges or contact with the underlying substrate. Our work provides an estimate of the thermal budget for controlled annealing of ferroelectric structures.

\section{\label{sec2:level1}Results}

\subsection{\label{sec2-1:level2}Extracting Twist and Strain}

The lowest-energy structure for bulk hBN exhibits an AA’ stacking conformation, shown in Fig. 1(a). Monolayer hBN lacks inversion symmetry due to the distinct B and N sublattices. In the AA’ stacking arrangement, the top layer is effectively rotated 60º (or equivalently, 180º) with respect to the bottom layer, which creates an inversion center at the interface. In antiparallel stacking, which mimics the bulk-like AA’ stacking with a small lateral displacement in the upper layer, the lattice will reconstruct to form primarily commensurate AA’ domains bounded by regions of energetically unfavorable AB’ stacking and incommensurate domain walls, shown in Fig. 1(d). The antiparallel configuration seen in Fig. 1(b) is not ferroelectric as it maintains inversion symmetry. However, in parallel stacking, the lattice reconstructs to the energetically favorable AB (BA) arrangement shown in Fig. 1(c). The AB and BA domains are separated by small regions of AA stacking and incommensurate domain walls as seen in Fig. 1(e). The parallel stacking configuration lacks inversion symmetry, allowing spontaneous polarization through a dipole moment from the distorted orbital of the nitrogen atom.\cite{Yasuda2021-cs}

\begin{figure*}
\includegraphics{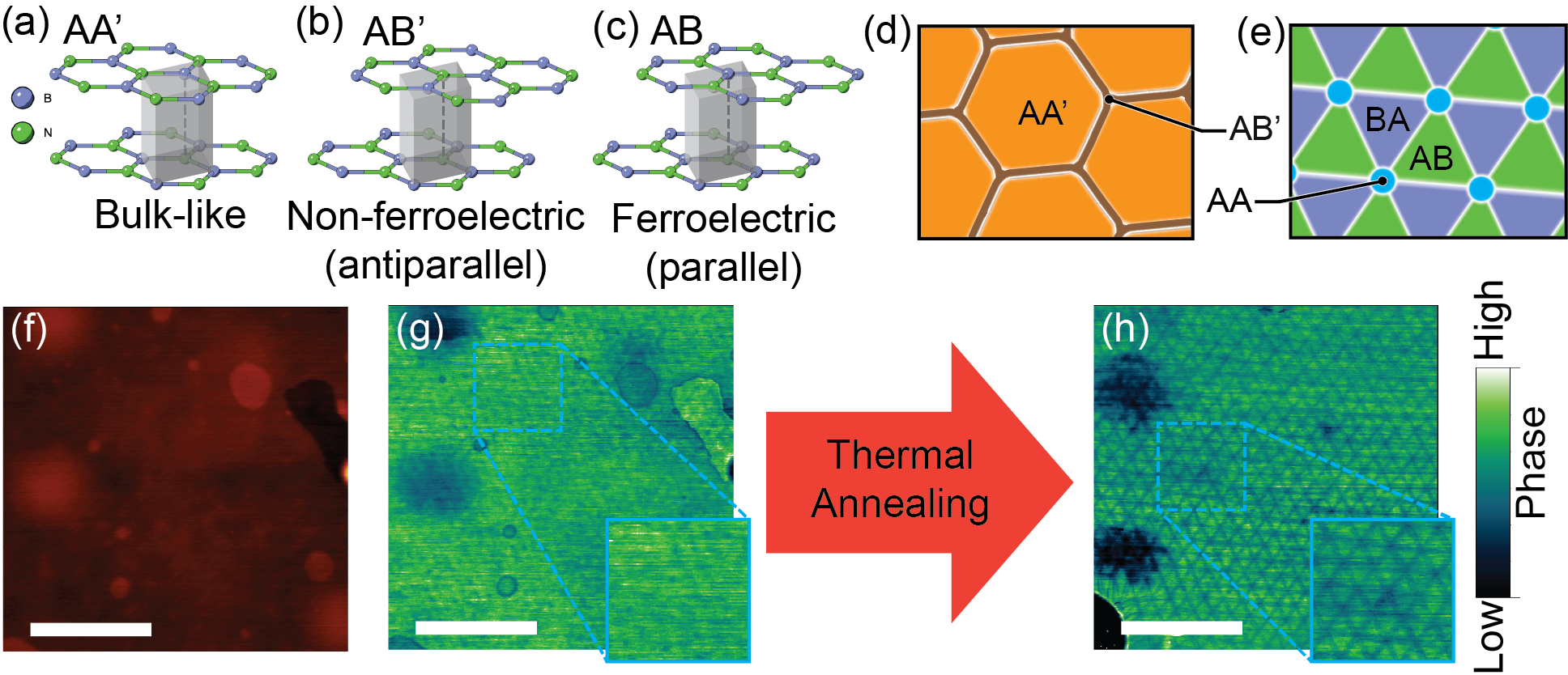}
\caption{\label{fig1:wide}} Crystallographic arrangement of hBN in its (a) bulk-like state as well as its (b) antiparallel and (c) parallel stacking configuration. After stacking, the moir{\'e} can relax into either (d) a bulk-like reconstruction with small AB’ regions or (c) a triangular reconstruction with oppositely polarized AB/BA domains, which is the ferroelectric reconstruction. In a ferroelectric hBN sample measured using contact resonance AFM (CR-AFM), the topography shows only morphological defects (f), while the phase channel shows the real-space moir{\'e} domains (g). These domains can be altered by thermal annealing as seen in the phase channel of the same region after treatment (h). All scale bars are 500 nm.
\end{figure*}

These reconstructions occur because the system seeks to minimize stacking fault energy by reducing the area of incommensurate regions at the competing energetic cost of mutual lattice adjustment (i.e., longitudinal and torsional strain). As the relative twist angle $\theta$ between the two layers approaches 0°, small amounts of uniaxial heterostrain $\epsilon$ introduce anisotropy into the reconstruction. The visual effect of strain on a lattice scales inversely with the twist angle ($\propto$1/$\theta$), which can magnify the effects of strain variation in marginally-twisted structures.\cite{Cosma2014-hl} 

The real-space domains and domain boundaries can be imaged using scanning probe techniques which are sensitive to the contrast in mechanical or electrical properties between the distinct regions produced by the reconstruction. These may include piezoresponse force microscopy (PFM),\cite{Yasuda2021-cs, Bai2020-wr} Kelvin probe force microscopy (KPFM),\cite{Vizner_Stern2021-gw, Woods2021-rz} contact resonance atomic force microscopy (CR-AFM),\cite{Ma2017-tx, Van_Es2018-lc} and other modes to measure nanomechanical response.\cite{Woods2014-bf} For many techniques measuring variation in mechanical properties, the phase (or frequency, if using a phase-locked loop) channel is most sensitive to these deviations.\cite{Ma2017-tx, Tamayo1997-to} While the topography for these samples only shows morphological defects as in Fig. 1(f), the CR-AFM phase channel (Fig. 1(g)) shows the domain walls of a ferroelectric reconstruction. Thermal annealing significantly alters the geometry of the reconstruction by enlarging and – in some cases – elongating the domains in Fig. 1(h).

Extracting twist and strain from real-space moir{\'e} lattices can be difficult, particularly from non-uniform or small twist angle moir{\'e}s. Commonly, real-space images are converted to a frequency-space representation via a fast Fourier transform (FFT), enabling the extraction of three reciprocal lattice vectors. In the limit of small deformations, assuming that biaxial strain and homostrain are negligible, these vectors can be fit to determine the wavelengths for a superlattice with small interfacial twist $\theta$ and uniaxial heterostrain with magnitude $\epsilon$ and direction $\phi$:\cite{Wang2023-ns}

\begin{eqnarray}
{\bf g}_{i=1,2} = \mathcal{E}^{\top}{\bf G}_{i=1,2}
\label{eq:one}.
\end{eqnarray}

\begin{eqnarray}
\mathcal{E} \equiv \mathcal{T}(\theta)+\mathcal{S}(\epsilon, \phi)
\label{eq:two},
\end{eqnarray}

\begin{eqnarray}
\mathcal{T} = 
\left(
\begin{array}{cc}
0 & -\theta \\
\theta & 0
\end{array}\right),
\mathcal{S}(\epsilon, \phi) = 
R^{\top}_{\phi}
\left(
\begin{array}{cc}
-\epsilon & 0\\
0 & \nu\epsilon
\end{array}\right)
R_{\phi}
\label{eq:three},
\end{eqnarray}

Where ${\bf g}_{i}$ and ${\bf G}_{i}$ are the moir{\'e} and undeformed reciprocal space lattice vectors for lattice vector $i$, $\mathcal{E}$ is the deformation lattice, $\mathcal{T}$ describes the deformation of the lattice due to twist, $\mathcal{S}$ describes the deformation of the lattice due to strain, $R_{\phi}$ is the two-dimensional rotation matrix using the shear angle $\phi$, and $\nu$ is Poisson’s ratio $\approx$ 0.21.\cite{Peng2012-zs} 

Following this set of equations, increasing the interfacial twist decreases the average size of the moir{\'e} domain, whereas strain contributes to the anisotropy of the domain in a direction determined by the shear angle. At low twist angles, the individual layers’ lattices relax so that the majority of each domain is commensurate. However, we retain a value of $\theta$ to facilitate comparison to other works.

\begin{figure*}
\includegraphics{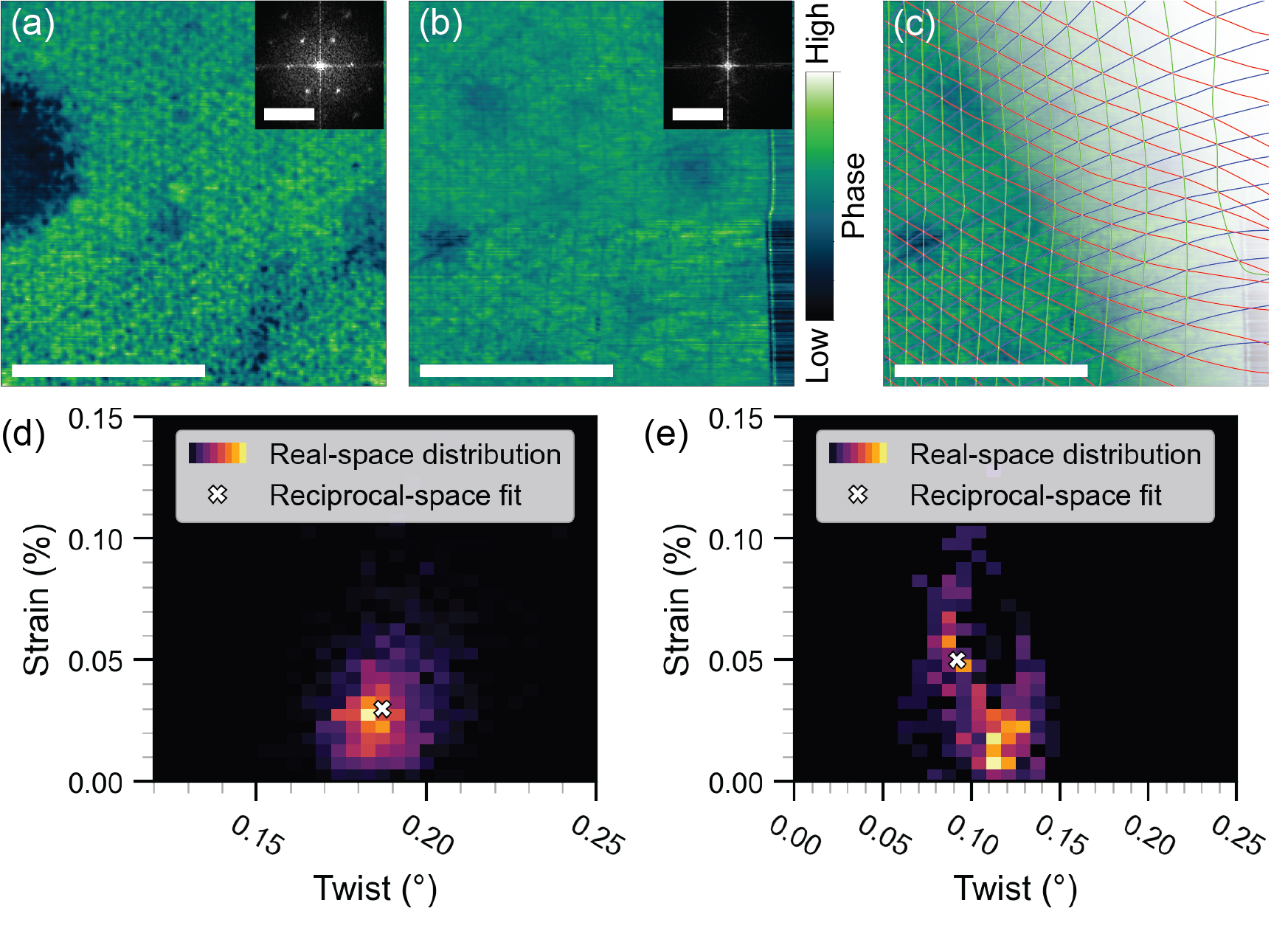}
\caption{\label{fig2:wide}} Analysis methods for near-zero twisted superlattices. The CR-AFM phase and related inset FFT are shown in (a) for a small, uniform moir{\'e} and (b) for a larger, non-uniform moir{\'e}. The domain walls of these near-zero twist superlattices can be manually traced in a vector program with an example seen in (c), and the twist and strain are computed on a per-triangle basis. For the uniform moir{\'e} from (a), the triangle-wise distribution (real-space distribution) of twist and strain is consistent with the peak twist and strain calculated directly from the FFT (reciprocal-space fit) in (d). For the non-uniform moir{\'e} from (b), the triangle-wise distribution and the peak values directly from the FFT are not well aligned in (e). Both 2D histograms in (d) and (e) are weighted with respect to area fraction of individual triangles. Scale bars on real-space images are 1 µm, and scale bars on FFT images are 25 µm$^{-1}$. 
\end{figure*}

For moir{\'e}s in which the periodicity is consistent throughout the field of view, as in Fig. 2(a), an FFT-based twist and strain extraction method can extract accurately representative values. However, in cases where the twist and/or strain exhibits only a few spatial periods or is heterogeneous within the imaged region, the moir{\'e} lacks long-range order needed to generate a strong peak in frequency space, and the FFT is weak or distorted as in Fig. 2(b). 

For these cases in which the frequency-space method is unsuitable, a real-space representation can provide complementary and precise representation of the moir{\'e}. The real-space moir{\'e} is extracted through identification of the individual moir{\'e} domains, by either automated methods\cite{Kazmierczak2021-lm, Alden2013-mi} or careful manual tracing of the domain boundaries\cite{Engelke2023-ez} as in Fig. 2(c). Each node and its six nearest neighbors can be found, and the real-space moir{\'e} lattice wavelengths $\lambda_{1}$, $\lambda_{2}$, and $\lambda_{3}$ can then be measured (neglecting curvature of the domain walls) for each individual triangle which comprises the relaxed half-moir{\'e} unit cell. Under the assumption that strain is purely uniaxial heterostrain, the per-triangle wavelengths can be transformed to reciprocal space and fit to equations (\ref{eq:one},~\ref{eq:two},~\ref{eq:three}). For relatively uniform regions as in Fig. 2(a), the real-space fit and the FFT fit show comparable results with respect to twist angle and strain, seen in  Fig. 2(d), although the real-space fit gives a more complete representation of the distribution.  For regions with stronger variation as in Fig. 2(b), the FFT has increasingly spread out and less intense peaks, so the moir{\'e} cannot be well-described by a single peak in frequency space, as seen in Fig. 2(e). 

\subsection{\label{sec2-2:level2}Temperature- and time-dependence of relaxation}

Studies have shown complete relaxation towards 0° of ferroelectric transition-metal dichalcogenide (TMDC) heterostructures assembled at an arbitrary angle upon heating to above 800°C,\cite{Baek2023-ic} suggesting that full relaxation is possible for other 2D ferroelectric structures, but the associated relaxation dynamics have not been characterized. To address this, we annealed ferroelectric hBN samples under conditions typical of assembly and sample processing. To mimic sample assembly in air, hBN/hBN samples were annealed on a hot plate in air for discrete time steps, then the moir{\'e}s were measured using PFM or CR-AFM in roughly the same position indicated in Fig. 3(a) and Fig. 3(d) each time.  We employed PFM before the utility of CR-AFM was realized, but the techniques are equivalent as they are both able to extract the spatial distribution of the moir{\'e} domain walls. The sample annealed at 170°C shows minimal changes to the domain size and isotropy between the as-fabricated sample in Fig. 3(b) and the same location in Fig. 3(c) after annealing for 20 min. In contrast, after 15 minutes of annealing at 400°C in air, a moir{\'e} in a different sample shows significant increase in domain size (Fig. 3(e) before annealing, Fig. 3(f) after.) 

\begin{figure*}
\includegraphics{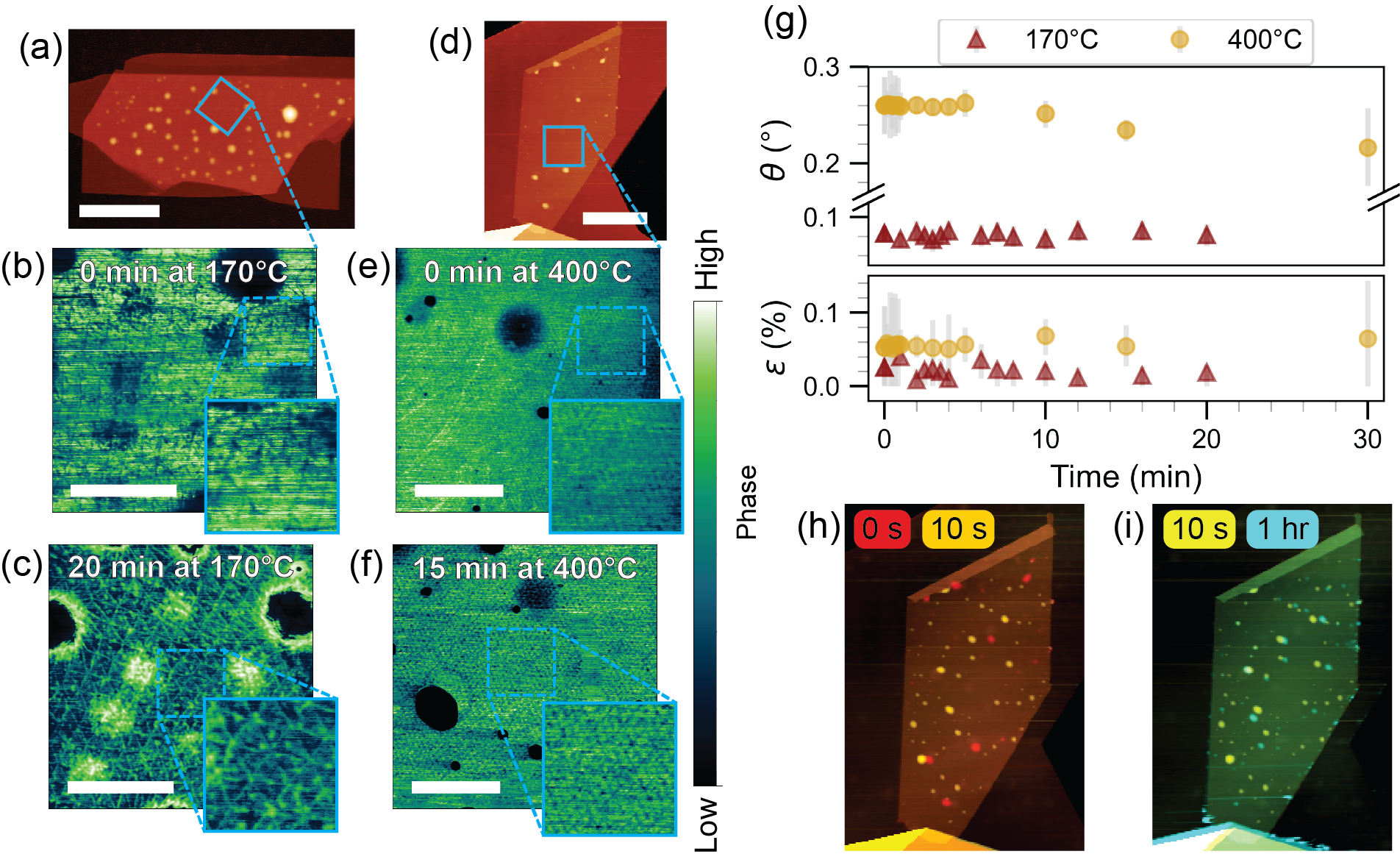}
\caption{\label{fig3:wide}} Two ferroelectric samples were annealed on a hot plate in air, and their moir{\'e}s were measured at discrete intervals using AFM techniques. An as-fabricated topographical map in (a) shows the measured area for the sample annealed at 170°C. From the PFM phase, the moir{\'e} before annealing (b) shows minimal differences with the moir{\'e} after 20 minutes at 170°C (c), conditions akin to heterostructure assembly. A second as-fabricated topographical map in (d) shows the measured area for the sample annealed at 400°C. In contrast to the case of 170°C annealing, there is a visible difference between the domains before annealing (e) and after 15 minutes at 400°C (f), this time mapped using CR-AFM phase instead of PFM phase. Using the direct FFT method, the twist and strain for each sample are shown over time in (g). Error bars were found as described in supplementary text S2. AFM topographical maps for the sample annealed at 400°C were overlaid to show bubble migration between (h) 0 s and 10 s as well as (i) 10 s and 1 hr.  Scale bars on images (a,d) are 5 µm and on images (b,c,e,f) are 1 µm. 
\end{figure*}

The uniform moir{\'e}s enabled unaxial heterostrain and twist extraction using the frequency-space method for all time points, shown in Fig. 3(g). The sample annealed at 170°C shows negligible change over the time period studied in terms of both moir{\'e} and topography aside from some bubble movement during the first minute (see Fig. S1). This suggests that assembly-like thermal conditions do not significantly affect the moir{\'e} geometry of ferroelectric hBN interfaces. The other sample annealed at 400°C, however, exhibited a decrease in twist. Previous works have shown that hBN encapsulation suppresses out-of-plane corrugations and supports interfacial reconstruction,\cite{Van_Winkle2023-ug, Baek2023-ic} which suggests that a larger number of hBN layers above the interface of interest should facilitate increased rotation as in-plane forces dominate. The sample annealed at 170°C should have experienced lower out-of-plane corrugations and therefore an energetic advantage to relax due to its thicker top hBN (10.2 nm), but its superlattice exhibited negligible changes compared to the sample annealed at 400°C, which had a thinner top hBN (3.3 nm). This indicates that the additional thermal energy is necessary to promote further relaxation. 

Comparison between the topography of the sample before annealing and after 10 s annealing at 400°C (Fig. 3(h)) reveals significant lateral bubble migration. However, comparing the topography at 10 s and 1 hr of annealing (Fig. 3(i)) the bubble movement is less evident. There is no moir{\'e} pattern evident in the bubbles seen in the phase maps in Fig. 3(e,f), which suggests that these bubbles are present at the interface between the flakes rather than at the interface between the bottom flake and underlying substrate and therefore may be more likely to influence the moir{\'e} relaxation behavior. In contrast, the measured twist angle of the sample relaxes on a different time scale, showing minimal change during the first few minutes followed by significant modification during the subsequent 25 minutes. This suggests that bubble migration is not the sole factor in reconstruction during thermal annealing.

\begin{figure*}
\includegraphics{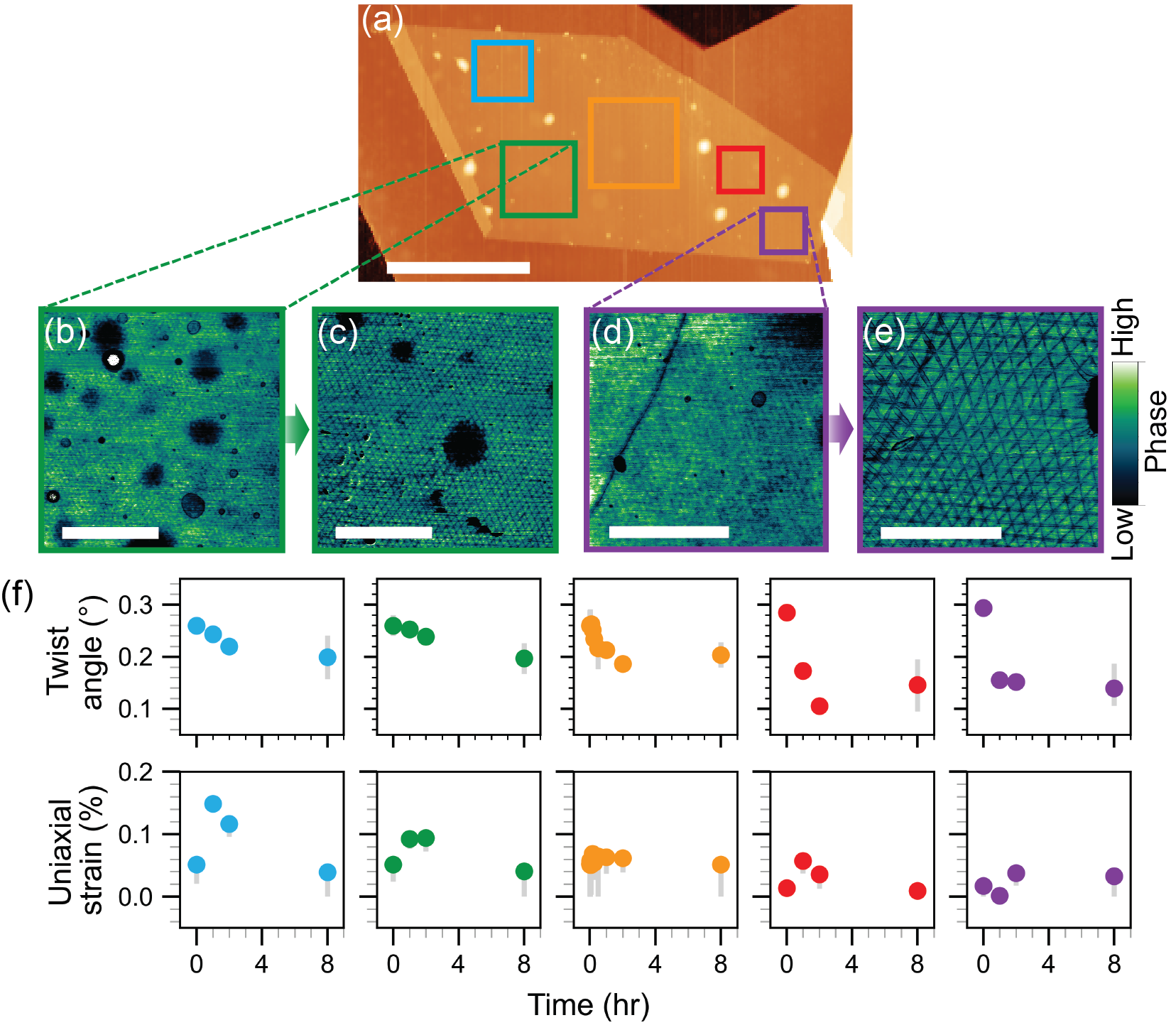}
\caption{\label{fig4:wide}} Various positions around the sample were measured as indicated in topography of the as-assembled structure in (a). Representative CR-AFM phase channel images from one position (b) before annealing and (c) after 8 hours shows a substantial increase in moir{\'e} period. The real-space moir{\'e} from the position closest to the step edge of the flake (d) before annealing and (e) after 8 hours shows a substantial increase in the unit cell size as well as a subtle increase in uniaxial strain compared to the pre-annealed sample. Local relaxation in twist and strain after annealing in air at 400°C is shown in (f) corresponding to the positions from the AFM topography in (a). Error bars were calculated as described in supplementary text S2. Scale bars are 5 µm in (a), 1 µm in (b,c), and 500 nm in (d,e). 
\end{figure*}

Time-dependent observations of annealed samples across a wider length scale (Fig. 4(a)) reveal that the evolution of thermal relaxation is strongly impacted by the local energy landscape. Substantial domain enlargement occurs between the as-assembled (Fig. 4(b)) and 8 hour time point (Fig. 4(c)). Continuing to anneal the sample in air at 400°C over several hours reveals two trends: a strong decrease and near-plateau in twist angle reduction, and a weak initial increase followed by a non-linear decrease in uniaxial strain highlighted in Fig. 4(f). The position closest to a large step shows an initial decrease in strain before increasing after a cumulative two hour annealing period. There is also slightly more strain after 8 hours annealing, which can be seen in comparing the anisotropy of the real-space moir{\'e} before annealing (Fig. 4(d)) and after 8 hours at 400°C (Fig. 4(e)), though this effect is magnified due to the reduction in twist angle. For all regions, despite sometimes substantial changes to the strain within the first two hours of annealing, the final strain after 8 hours only shows minor changes compared to the initial as-assembled strain.

\begin{figure*}
\includegraphics{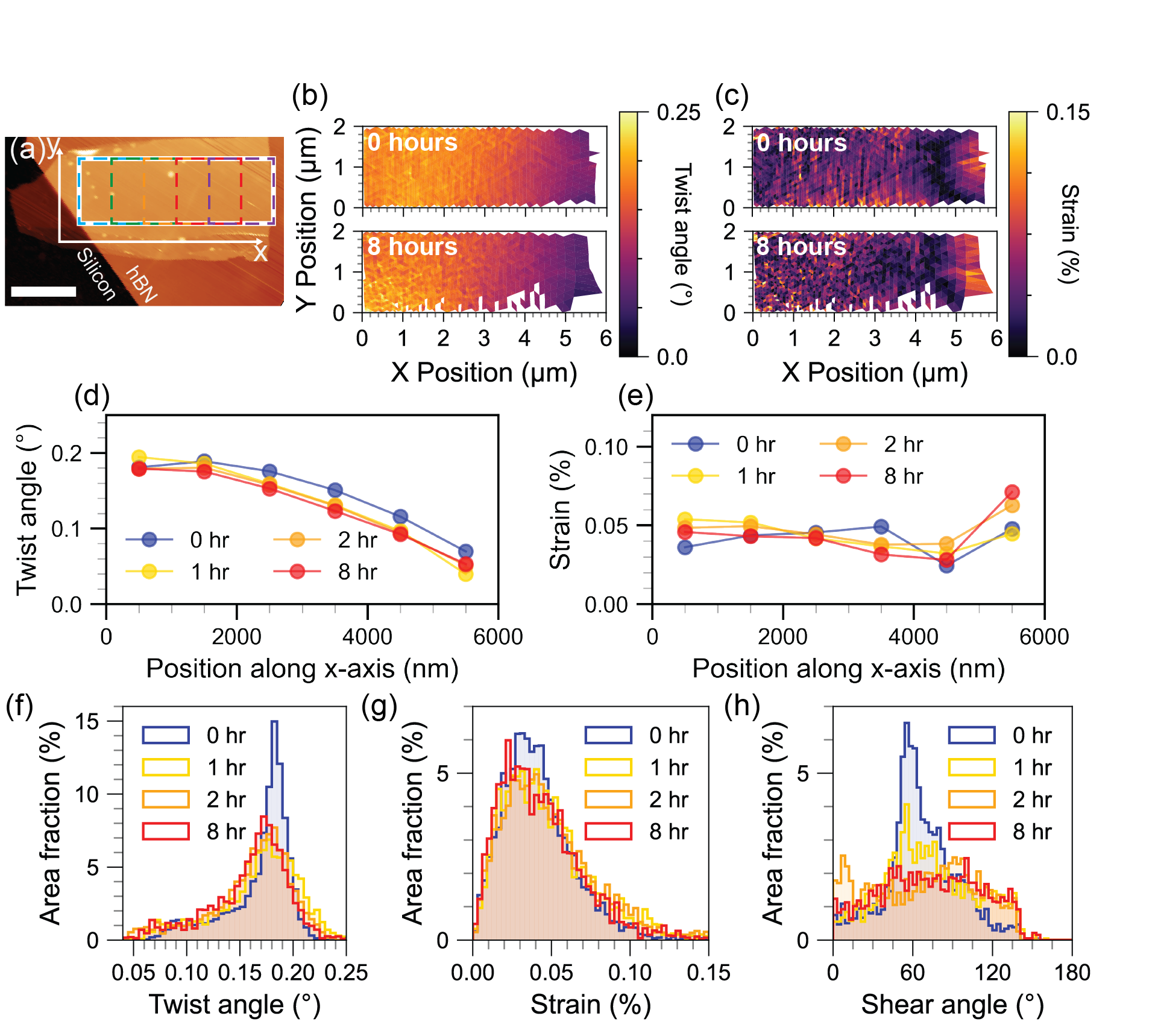}
\caption{\label{fig5:wide}} (a) AFM topography of a sample annealed in vacuum (p $\leq$ 5E-6 Torr) at 400°C showing several overlapping positions where the superlattice was measured. The overlapping positions were stitched together to get a continuous map across the sample. The corresponding (b) twist and (c) strain maps for the as-assembled moir{\'e} and the annealed moir{\'e} are shown for each individual half-moir{\'e} unit cell. The spatially-averaged twist and strain along the x-axis are shown in in (d) and (e), respectively. The distributions of (f) twist, (g) strain, and (h) shear angle for the sample at several time points were extracted from these maps. Scale bars are 2 µm.
\end{figure*}

Contact between the relaxing top flake and the underlying bulk substrate affects relaxation behavior as well. A sample with a fairly large and non-uniform lattice, requiring the real-space extraction method, was annealed at 400°C in vacuum for several hours. The real-space vector mapping method was applied as there was a large degree of inhomogeneity across the sample after assembly. The strain and twist can be measured for each half of the unit cell (i.e., each triangle), and the evolution in the distribution of the twist angle and strain can be examined. If the strain and twist for each discrete annealing step are plotted as a function of position within the sample as in Fig. 5(b-c), there is a distinct gradient within the sample. When looking at the spatially-averaged twist along the x-axis in Fig. 5(f), it becomes clear that the interfacial twist is pinned in the regions closest to the silicon/hBN contact region. The spatially-averaged strain along the x-axis in Fig. 5(f) shows pinning within the center of the top flake and then variations towards both ends, though the end furthest from the silicon/hBN contact region shows the largest change. Similarly to the sample annealed in-air, the twist angle decreases in Fig. 5(d), and the uniaxial strain initially increases slightly, then decreases in Fig. 5(g). The width of the twist angle distribution is small before the anneal and broadens with one hour of annealing, though it does not continue to increase. This suggests that annealing perturbs the uniformity of the interfacial twist, corroborating previous work on marginally twisted heterostructures,\cite{Woods2016-lb} though the amount of perturbation is not monotonic with the duration of annealing. While the width of the strain distribution does not change significantly over time, the width of the shear angle distribution in Fig. 5(h) increases. This implies that the uniformity of the strain directionality changes while the uniformity of the strain magnitude itself does not.

\section{\label{sec3:level1}Discussion}

Thermal relaxation of moir{\'e} reconstructions can be challenging to untangle. Mechanisms governing a finite rate of relaxation are difficult to discern. In particular, strain has been shown to vary substantially across a small region even when interfacial twist does not vary substantially,\cite{Kazmierczak2021-lm, De_Jong2022-fo} making its investigation and control even more significant. Our work has shown a reduction in twist angle that saturates at long annealing times as well as an increase in strain at short annealing times. Both parameters become less spatially-uniform compared to the moir{\'e} before annealing. The driving force behind these trends is unclear. 

Previous works suggest that thermal relaxation is primarily manifested through local atomic rearrangements rather than macroscopic flake rotation.\cite{Baek2023-ic, Alden2013-mi} In encapsulated TMDC homobilayers, these rearrangements occur non-preferentially within both the top and bottom layer,\cite{Baek2023-ic} likely due to large lattice mismatch at the TMDC and hBN interfaces.\cite{Liao2022-zl} Non-encapsulated TMDC homobilayers show reduced or no further in-plane reconstructions at the same temperatures at which a fully-encapsulated sample reaches complete commensuration,\cite{Baek2023-ic} presumably arising from competition with out-of-plane corrugations that are otherwise suppressed during encapsulation.\cite{Van_Winkle2023-ug} Local rearrangement may explain the minor absolute changes in strain before and after several hours of annealing seen in Fig. 4(f). However, macroscopic rotation of entire flakes after thermal annealing has also been observed optically in the case where the top layer is situated entirely on a thicker bottom flake without overhang onto a rough underlying substrate, shown in Fig S3 as well as previous works.\cite{Woods2016-lb, Wang2015-ly, Wang2016-yj} This suggests that increased friction due to layer conformation with a rough substrate\cite{Dong2014-iu} is not negligible. It may explain the local fluctuations in strain seen in Fig. 5(e) as the atoms in the portion of the top flake furthest from the underlying substrate could more easily rearrange compared to the regions closer to contact with the silicon.

Other frictional effects must also be considered during thermal relaxation. In systems with small or nonexistent lattice mismatch, rotational friction peaks at commensurate states,\cite{Dienwiebel2004-da} which would seemingly account for a plateau in twist angle reduction as subsequently larger portions of the sample achieve commensuration through reconstruction. There is also evidence for a strain-dependent reduction in friction.\cite{Zhang2019-fb} As the strain is reduced, the friction may increase as there is a lower energetic driving force for rearrangement, potentially contributing to the plateau in twist angle reduction. The relaxation rate during annealing was anticipated to be faster because the coefficient of friction between two interfaces is reduced in vacuum as there are fewer adsorbates.\cite{Martin1993-ge, He1999-ed} While this suggests that other frictional effects may be dominating, it also offers an opportunity for future works to study the effect of annealing environment on assembled heterostructures. 

Metastable rotational states may also contribute to a relaxation plateau. Theory indicates that there is a local energetic minima for parallel-stacked AB/BA domains near zero interfacial twist, and this local minima shifts closer to zero as the size of the flake increases.\cite{Zhu2019-ay, Bagchi2020-jm} Experimental work with graphene/hBN supports this with a metastable state close to 0.7°,\cite{Woods2016-lb} though the effect is not consistent, and it has not been shown in homobilayers. Further work suggests that heterostrain can stabilize the interfacial twist above zero even at elevated temperatures,\cite{Yang2023-qf} providing another factor that may contribute to a metastable state.

Differences in relaxation rates may be partially attributed to interfacial contamination. Trapped water or alcohol molecules have been shown to increase the interfacial friction, and the increase scales proportionally with the amount of solvent.\cite{Lee2017-xn, Dollekamp2019-fh} Previous work shows that significant rotation does not always occur in heterostructures, particularly those with large amounts of bubbles or other interlayer contaminants.\cite{Woods2016-lb, Wang2015-ly} This suggests that twist angle reduction in heterostructures with larger amounts of interfacial contamination would plateau earlier than in clean heterostructures and offers an opportunity for further investigation of relaxation rate with respect to sample cleanliness.

\section{\label{sec4:level1}Conclusions}

We have shown evidence of a finite, non-linear twist and strain relaxation rate in parallel-stacked hBN structures through thermal annealing. Successive annealing steps reduce the interfacial twist to a saturation point as well as reducing twist uniformity. Uniaxial heterostrain tends towards an initial increase at short annealing times and then decreases after further annealing, though its directionality decreases in uniformity. However, through considerations during assembly, it is possible to exploit pinning points to create an intentional and directional gradient in twist and strain without mechanical interference. This work provides a greater understanding of the effect of annealing on near-zero moir{\'e} superlattices and other steps towards controlling moir{\'e} geometry after fabrication.

\section*{\label{sec5:level1}Supplementary material}

See the supplementary material for detailed methods, topography scans of the samples annealed in air at 170°C and 400°C, and optical images of hBN/hBN samples showing macroscopic rotation after annealing in air at 400°C. 

\begin{acknowledgments}

The authors thank Chaitrali Duse and Marc Kastner for fruitful discussions. Annealing procedures and moir{\'e} measurements were supported by the US Department of Energy, Office of Science, Basic Energy Sciences, Materials Sciences and Engineering Division, under Contract DE-AC02-76SF00515. Development of tools for robotic stacking of 2D materials were supported by SLAC National Accelerator Laboratory under the Q-BALMS Laboratory Directed Research and Development funds. Initial development of twisted hBN fabrication protocols was supported by supported the US Department of Energy (DOE) Basic Energy Sciences grant DE-SC0021984. The process for fabricating polymeric stamps was supported by National Science Foundation DMR- 2018008. Part of this work was performed at the Stanford Nano Shared Facilities (SNSF), supported by the National Science Foundation under award ECCS-2026822, as well as in the nano@Stanford labs, which are supported by the National Science Foundation as part of the National Nanotechnology Coordinated Infrastructure under award ECCS-1542152. M.H. acknowledges partial support from the US Department of Defense through the Graduate Fellowship in STEM Diversity program. 

\end{acknowledgments}

\section*{\label{sec6:level1}Author declarations}
\subsection*{\label{sec6-1:level2}Conflicts of Interest}
The authors have no conflicts to disclose.
\subsection*{\label{sec6-2:level2}Author Contributions}

\textbf{M. Hocking:} Conceptualization (equal); Investigation (lead); Formal analysis (lead); Software (lead); Visualization (lead); Writing – original draft (lead); Writing – review $\&$ editing (equal). \textbf{C.E. Henzinger:} Formal analysis (support). \textbf{S. Tran:} Resources (equal). \textbf{M. Pendharkar:} Resources (equal). \textbf{N.J. Bittner:} Resources (equal). \textbf{T. Taniguchi:} Resources (equal). \textbf{K. Watanabe:} Resources (equal). \textbf{D. Goldhaber-Gordon:} Writing – review $\&$ editing (equal); Supervision (equal); Funding acquisition (equal). \textbf{A.J. Mannix:} Conceptualization (equal); Writing – original draft (support); Writing – review $\&$ editing (equal); Supervision (equal); Funding acquisition (equal).

\subsection*{\label{sec6-3:level2}Data availability}
The data that supports the findings of this study are available from the corresponding author upon reasonable request. 

\section*{\label{sec7:leve1}References}
\bibliography{references}

\begin{thebibliography}{46}%
\makeatletter
\providecommand \@ifxundefined [1]{%
 \@ifx{#1\undefined}
}%
\providecommand \@ifnum [1]{%
 \ifnum #1\expandafter \@firstoftwo
 \else \expandafter \@secondoftwo
 \fi
}%
\providecommand \@ifx [1]{%
 \ifx #1\expandafter \@firstoftwo
 \else \expandafter \@secondoftwo
 \fi
}%
\providecommand \natexlab [1]{#1}%
\providecommand \enquote  [1]{``#1''}%
\providecommand \bibnamefont  [1]{#1}%
\providecommand \bibfnamefont [1]{#1}%
\providecommand \citenamefont [1]{#1}%
\providecommand \href@noop [0]{\@secondoftwo}%
\providecommand \href [0]{\begingroup \@sanitize@url \@href}%
\providecommand \@href[1]{\@@startlink{#1}\@@href}%
\providecommand \@@href[1]{\endgroup#1\@@endlink}%
\providecommand \@sanitize@url [0]{\catcode `\\12\catcode `\$12\catcode `\&12\catcode `\#12\catcode `\^12\catcode `\_12\catcode `\%12\relax}%
\providecommand \@@startlink[1]{}%
\providecommand \@@endlink[0]{}%
\providecommand \url  [0]{\begingroup\@sanitize@url \@url }%
\providecommand \@url [1]{\endgroup\@href {#1}{\urlprefix }}%
\providecommand \urlprefix  [0]{URL }%
\providecommand \Eprint [0]{\href }%
\providecommand \doibase [0]{http://dx.doi.org/}%
\providecommand \selectlanguage [0]{\@gobble}%
\providecommand \bibinfo  [0]{\@secondoftwo}%
\providecommand \bibfield  [0]{\@secondoftwo}%
\providecommand \translation [1]{[#1]}%
\providecommand \BibitemOpen [0]{}%
\providecommand \bibitemStop [0]{}%
\providecommand \bibitemNoStop [0]{.\EOS\space}%
\providecommand \EOS [0]{\spacefactor3000\relax}%
\providecommand \BibitemShut  [1]{\csname bibitem#1\endcsname}%
\let\auto@bib@innerbib\@empty
\bibitem [{\citenamefont {Cao}\ \emph {et~al.}(2018)\citenamefont {Cao}, \citenamefont {Fatemi}, \citenamefont {Fang}, \citenamefont {Watanabe}, \citenamefont {Taniguchi}, \citenamefont {Kaxiras},\ and\ \citenamefont {Jarillo-Herrero}}]{Cao2018-sl}%
  \BibitemOpen
  \bibfield  {author} {\bibinfo {author} {\bibfnamefont {Y.}~\bibnamefont {Cao}}, \bibinfo {author} {\bibfnamefont {V.}~\bibnamefont {Fatemi}}, \bibinfo {author} {\bibfnamefont {S.}~\bibnamefont {Fang}}, \bibinfo {author} {\bibfnamefont {K.}~\bibnamefont {Watanabe}}, \bibinfo {author} {\bibfnamefont {T.}~\bibnamefont {Taniguchi}}, \bibinfo {author} {\bibfnamefont {E.}~\bibnamefont {Kaxiras}}, \ and\ \bibinfo {author} {\bibfnamefont {P.}~\bibnamefont {Jarillo-Herrero}},\ }\href@noop {} {\bibfield  {journal} {\bibinfo  {journal} {Nature}\ }\textbf {\bibinfo {volume} {556}},\ \bibinfo {pages} {43} (\bibinfo {year} {2018})}\BibitemShut {NoStop}%
\bibitem [{\citenamefont {Sharpe}\ \emph {et~al.}(2019)\citenamefont {Sharpe}, \citenamefont {Fox}, \citenamefont {Barnard}, \citenamefont {Finney}, \citenamefont {Watanabe}, \citenamefont {Taniguchi}, \citenamefont {Kastner},\ and\ \citenamefont {Goldhaber-Gordon}}]{Sharpe2019-ot}%
  \BibitemOpen
  \bibfield  {author} {\bibinfo {author} {\bibfnamefont {A.~L.}\ \bibnamefont {Sharpe}}, \bibinfo {author} {\bibfnamefont {E.~J.}\ \bibnamefont {Fox}}, \bibinfo {author} {\bibfnamefont {A.~W.}\ \bibnamefont {Barnard}}, \bibinfo {author} {\bibfnamefont {J.}~\bibnamefont {Finney}}, \bibinfo {author} {\bibfnamefont {K.}~\bibnamefont {Watanabe}}, \bibinfo {author} {\bibfnamefont {T.}~\bibnamefont {Taniguchi}}, \bibinfo {author} {\bibfnamefont {M.~A.}\ \bibnamefont {Kastner}}, \ and\ \bibinfo {author} {\bibfnamefont {D.}~\bibnamefont {Goldhaber-Gordon}},\ }\href@noop {} {\bibfield  {journal} {\bibinfo  {journal} {Science}\ }\textbf {\bibinfo {volume} {365}},\ \bibinfo {pages} {605} (\bibinfo {year} {2019})}\BibitemShut {NoStop}%
\bibitem [{\citenamefont {Regan}\ \emph {et~al.}(2020)\citenamefont {Regan}, \citenamefont {Wang}, \citenamefont {Jin}, \citenamefont {Bakti~Utama}, \citenamefont {Gao}, \citenamefont {Wei}, \citenamefont {Zhao}, \citenamefont {Zhao}, \citenamefont {Zhang}, \citenamefont {Yumigeta}, \citenamefont {Blei}, \citenamefont {Carlstr{\"o}m}, \citenamefont {Watanabe}, \citenamefont {Taniguchi}, \citenamefont {Tongay}, \citenamefont {Crommie}, \citenamefont {Zettl},\ and\ \citenamefont {Wang}}]{Regan2020-pd}%
  \BibitemOpen
  \bibfield  {author} {\bibinfo {author} {\bibfnamefont {E.~C.}\ \bibnamefont {Regan}}, \bibinfo {author} {\bibfnamefont {D.}~\bibnamefont {Wang}}, \bibinfo {author} {\bibfnamefont {C.}~\bibnamefont {Jin}}, \bibinfo {author} {\bibfnamefont {M.~I.}\ \bibnamefont {Bakti~Utama}}, \bibinfo {author} {\bibfnamefont {B.}~\bibnamefont {Gao}}, \bibinfo {author} {\bibfnamefont {X.}~\bibnamefont {Wei}}, \bibinfo {author} {\bibfnamefont {S.}~\bibnamefont {Zhao}}, \bibinfo {author} {\bibfnamefont {W.}~\bibnamefont {Zhao}}, \bibinfo {author} {\bibfnamefont {Z.}~\bibnamefont {Zhang}}, \bibinfo {author} {\bibfnamefont {K.}~\bibnamefont {Yumigeta}}, \bibinfo {author} {\bibfnamefont {M.}~\bibnamefont {Blei}}, \bibinfo {author} {\bibfnamefont {J.~D.}\ \bibnamefont {Carlstr{\"o}m}}, \bibinfo {author} {\bibfnamefont {K.}~\bibnamefont {Watanabe}}, \bibinfo {author} {\bibfnamefont {T.}~\bibnamefont {Taniguchi}}, \bibinfo {author} {\bibfnamefont {S.}~\bibnamefont {Tongay}}, \bibinfo {author} {\bibfnamefont {M.}~\bibnamefont
  {Crommie}}, \bibinfo {author} {\bibfnamefont {A.}~\bibnamefont {Zettl}}, \ and\ \bibinfo {author} {\bibfnamefont {F.}~\bibnamefont {Wang}},\ }\href@noop {} {\bibfield  {journal} {\bibinfo  {journal} {Nature}\ }\textbf {\bibinfo {volume} {579}},\ \bibinfo {pages} {359} (\bibinfo {year} {2020})}\BibitemShut {NoStop}%
\bibitem [{\citenamefont {Ko}\ \emph {et~al.}(2023)\citenamefont {Ko}, \citenamefont {Yuk}, \citenamefont {Engelke}, \citenamefont {Carr}, \citenamefont {Kim}, \citenamefont {Park}, \citenamefont {Heo}, \citenamefont {Kim}, \citenamefont {Kim}, \citenamefont {Kim}, \citenamefont {Taniguchi}, \citenamefont {Watanabe}, \citenamefont {Park}, \citenamefont {Kaxiras}, \citenamefont {Yang}, \citenamefont {Kim},\ and\ \citenamefont {Yoo}}]{Ko2023-zt}%
  \BibitemOpen
  \bibfield  {author} {\bibinfo {author} {\bibfnamefont {K.}~\bibnamefont {Ko}}, \bibinfo {author} {\bibfnamefont {A.}~\bibnamefont {Yuk}}, \bibinfo {author} {\bibfnamefont {R.}~\bibnamefont {Engelke}}, \bibinfo {author} {\bibfnamefont {S.}~\bibnamefont {Carr}}, \bibinfo {author} {\bibfnamefont {J.}~\bibnamefont {Kim}}, \bibinfo {author} {\bibfnamefont {D.}~\bibnamefont {Park}}, \bibinfo {author} {\bibfnamefont {H.}~\bibnamefont {Heo}}, \bibinfo {author} {\bibfnamefont {H.-M.}\ \bibnamefont {Kim}}, \bibinfo {author} {\bibfnamefont {S.-G.}\ \bibnamefont {Kim}}, \bibinfo {author} {\bibfnamefont {H.}~\bibnamefont {Kim}}, \bibinfo {author} {\bibfnamefont {T.}~\bibnamefont {Taniguchi}}, \bibinfo {author} {\bibfnamefont {K.}~\bibnamefont {Watanabe}}, \bibinfo {author} {\bibfnamefont {H.}~\bibnamefont {Park}}, \bibinfo {author} {\bibfnamefont {E.}~\bibnamefont {Kaxiras}}, \bibinfo {author} {\bibfnamefont {S.~M.}\ \bibnamefont {Yang}}, \bibinfo {author} {\bibfnamefont {P.}~\bibnamefont {Kim}}, \ and\ \bibinfo
  {author} {\bibfnamefont {H.}~\bibnamefont {Yoo}},\ }\href@noop {} {\bibfield  {journal} {\bibinfo  {journal} {Nat. Mater.}\ } (\bibinfo {year} {2023})}\BibitemShut {NoStop}%
\bibitem [{\citenamefont {Deb}\ \emph {et~al.}(2022)\citenamefont {Deb}, \citenamefont {Cao}, \citenamefont {Raab}, \citenamefont {Watanabe}, \citenamefont {Taniguchi}, \citenamefont {Goldstein}, \citenamefont {Kronik}, \citenamefont {Urbakh}, \citenamefont {Hod},\ and\ \citenamefont {Ben~Shalom}}]{Deb2022-zw}%
  \BibitemOpen
  \bibfield  {author} {\bibinfo {author} {\bibfnamefont {S.}~\bibnamefont {Deb}}, \bibinfo {author} {\bibfnamefont {W.}~\bibnamefont {Cao}}, \bibinfo {author} {\bibfnamefont {N.}~\bibnamefont {Raab}}, \bibinfo {author} {\bibfnamefont {K.}~\bibnamefont {Watanabe}}, \bibinfo {author} {\bibfnamefont {T.}~\bibnamefont {Taniguchi}}, \bibinfo {author} {\bibfnamefont {M.}~\bibnamefont {Goldstein}}, \bibinfo {author} {\bibfnamefont {L.}~\bibnamefont {Kronik}}, \bibinfo {author} {\bibfnamefont {M.}~\bibnamefont {Urbakh}}, \bibinfo {author} {\bibfnamefont {O.}~\bibnamefont {Hod}}, \ and\ \bibinfo {author} {\bibfnamefont {M.}~\bibnamefont {Ben~Shalom}},\ }\href@noop {} {\bibfield  {journal} {\bibinfo  {journal} {Nature}\ } (\bibinfo {year} {2022})}\BibitemShut {NoStop}%
\bibitem [{\citenamefont {Weston}\ \emph {et~al.}(2022)\citenamefont {Weston}, \citenamefont {Castanon}, \citenamefont {Enaldiev}, \citenamefont {Ferreira}, \citenamefont {Bhattacharjee}, \citenamefont {Xu}, \citenamefont {Corte-Le{\'o}n}, \citenamefont {Wu}, \citenamefont {Clark}, \citenamefont {Summerfield}, \citenamefont {Hashimoto}, \citenamefont {Gao}, \citenamefont {Wang}, \citenamefont {Hamer}, \citenamefont {Read}, \citenamefont {Fumagalli}, \citenamefont {Kretinin}, \citenamefont {Haigh}, \citenamefont {Kazakova}, \citenamefont {Geim}, \citenamefont {Fal'ko},\ and\ \citenamefont {Gorbachev}}]{Weston2022-ju}%
  \BibitemOpen
  \bibfield  {author} {\bibinfo {author} {\bibfnamefont {A.}~\bibnamefont {Weston}}, \bibinfo {author} {\bibfnamefont {E.~G.}\ \bibnamefont {Castanon}}, \bibinfo {author} {\bibfnamefont {V.}~\bibnamefont {Enaldiev}}, \bibinfo {author} {\bibfnamefont {F.}~\bibnamefont {Ferreira}}, \bibinfo {author} {\bibfnamefont {S.}~\bibnamefont {Bhattacharjee}}, \bibinfo {author} {\bibfnamefont {S.}~\bibnamefont {Xu}}, \bibinfo {author} {\bibfnamefont {H.}~\bibnamefont {Corte-Le{\'o}n}}, \bibinfo {author} {\bibfnamefont {Z.}~\bibnamefont {Wu}}, \bibinfo {author} {\bibfnamefont {N.}~\bibnamefont {Clark}}, \bibinfo {author} {\bibfnamefont {A.}~\bibnamefont {Summerfield}}, \bibinfo {author} {\bibfnamefont {T.}~\bibnamefont {Hashimoto}}, \bibinfo {author} {\bibfnamefont {Y.}~\bibnamefont {Gao}}, \bibinfo {author} {\bibfnamefont {W.}~\bibnamefont {Wang}}, \bibinfo {author} {\bibfnamefont {M.}~\bibnamefont {Hamer}}, \bibinfo {author} {\bibfnamefont {H.}~\bibnamefont {Read}}, \bibinfo {author} {\bibfnamefont {L.}~\bibnamefont
  {Fumagalli}}, \bibinfo {author} {\bibfnamefont {A.~V.}\ \bibnamefont {Kretinin}}, \bibinfo {author} {\bibfnamefont {S.~J.}\ \bibnamefont {Haigh}}, \bibinfo {author} {\bibfnamefont {O.}~\bibnamefont {Kazakova}}, \bibinfo {author} {\bibfnamefont {A.~K.}\ \bibnamefont {Geim}}, \bibinfo {author} {\bibfnamefont {V.~I.}\ \bibnamefont {Fal'ko}}, \ and\ \bibinfo {author} {\bibfnamefont {R.}~\bibnamefont {Gorbachev}},\ }\href@noop {} {\bibfield  {journal} {\bibinfo  {journal} {Nat. Nanotechnol.}\ }\textbf {\bibinfo {volume} {17}},\ \bibinfo {pages} {390} (\bibinfo {year} {2022})}\BibitemShut {NoStop}%
\bibitem [{\citenamefont {Woods}\ \emph {et~al.}(2021)\citenamefont {Woods}, \citenamefont {Ares}, \citenamefont {Nevison-Andrews}, \citenamefont {Holwill}, \citenamefont {Fabregas}, \citenamefont {Guinea}, \citenamefont {Geim}, \citenamefont {Novoselov}, \citenamefont {Walet},\ and\ \citenamefont {Fumagalli}}]{Woods2021-rz}%
  \BibitemOpen
  \bibfield  {author} {\bibinfo {author} {\bibfnamefont {C.~R.}\ \bibnamefont {Woods}}, \bibinfo {author} {\bibfnamefont {P.}~\bibnamefont {Ares}}, \bibinfo {author} {\bibfnamefont {H.}~\bibnamefont {Nevison-Andrews}}, \bibinfo {author} {\bibfnamefont {M.~J.}\ \bibnamefont {Holwill}}, \bibinfo {author} {\bibfnamefont {R.}~\bibnamefont {Fabregas}}, \bibinfo {author} {\bibfnamefont {F.}~\bibnamefont {Guinea}}, \bibinfo {author} {\bibfnamefont {A.~K.}\ \bibnamefont {Geim}}, \bibinfo {author} {\bibfnamefont {K.~S.}\ \bibnamefont {Novoselov}}, \bibinfo {author} {\bibfnamefont {N.~R.}\ \bibnamefont {Walet}}, \ and\ \bibinfo {author} {\bibfnamefont {L.}~\bibnamefont {Fumagalli}},\ }\href@noop {} {\bibfield  {journal} {\bibinfo  {journal} {Nat. Commun.}\ }\textbf {\bibinfo {volume} {12}},\ \bibinfo {pages} {347} (\bibinfo {year} {2021})}\BibitemShut {NoStop}%
\bibitem [{\citenamefont {Yasuda}\ \emph {et~al.}(2021)\citenamefont {Yasuda}, \citenamefont {Wang}, \citenamefont {Watanabe}, \citenamefont {Taniguchi},\ and\ \citenamefont {Jarillo-Herrero}}]{Yasuda2021-cs}%
  \BibitemOpen
  \bibfield  {author} {\bibinfo {author} {\bibfnamefont {K.}~\bibnamefont {Yasuda}}, \bibinfo {author} {\bibfnamefont {X.}~\bibnamefont {Wang}}, \bibinfo {author} {\bibfnamefont {K.}~\bibnamefont {Watanabe}}, \bibinfo {author} {\bibfnamefont {T.}~\bibnamefont {Taniguchi}}, \ and\ \bibinfo {author} {\bibfnamefont {P.}~\bibnamefont {Jarillo-Herrero}},\ }\href@noop {} {\bibfield  {journal} {\bibinfo  {journal} {Science}\ } (\bibinfo {year} {2021})}\BibitemShut {NoStop}%
\bibitem [{\citenamefont {Vizner~Stern}\ \emph {et~al.}(2021)\citenamefont {Vizner~Stern}, \citenamefont {Waschitz}, \citenamefont {Cao}, \citenamefont {Nevo}, \citenamefont {Watanabe}, \citenamefont {Taniguchi}, \citenamefont {Sela}, \citenamefont {Urbakh}, \citenamefont {Hod},\ and\ \citenamefont {Ben~Shalom}}]{Vizner_Stern2021-gw}%
  \BibitemOpen
  \bibfield  {author} {\bibinfo {author} {\bibfnamefont {M.}~\bibnamefont {Vizner~Stern}}, \bibinfo {author} {\bibfnamefont {Y.}~\bibnamefont {Waschitz}}, \bibinfo {author} {\bibfnamefont {W.}~\bibnamefont {Cao}}, \bibinfo {author} {\bibfnamefont {I.}~\bibnamefont {Nevo}}, \bibinfo {author} {\bibfnamefont {K.}~\bibnamefont {Watanabe}}, \bibinfo {author} {\bibfnamefont {T.}~\bibnamefont {Taniguchi}}, \bibinfo {author} {\bibfnamefont {E.}~\bibnamefont {Sela}}, \bibinfo {author} {\bibfnamefont {M.}~\bibnamefont {Urbakh}}, \bibinfo {author} {\bibfnamefont {O.}~\bibnamefont {Hod}}, \ and\ \bibinfo {author} {\bibfnamefont {M.}~\bibnamefont {Ben~Shalom}},\ }\href@noop {} {\bibfield  {journal} {\bibinfo  {journal} {Science}\ } (\bibinfo {year} {2021})}\BibitemShut {NoStop}%
\bibitem [{\citenamefont {Frauni{\'e}}\ \emph {et~al.}(2023)\citenamefont {Frauni{\'e}}, \citenamefont {Jamil}, \citenamefont {Kantelberg}, \citenamefont {Roux}, \citenamefont {Petit}, \citenamefont {Lepleux}, \citenamefont {Pacheco}, \citenamefont {Watanabe}, \citenamefont {Taniguchi}, \citenamefont {Jacques}, \citenamefont {Lombez}, \citenamefont {Glazov}, \citenamefont {Lassagne}, \citenamefont {Marie},\ and\ \citenamefont {Robert}}]{Fraunie2023-gj}%
  \BibitemOpen
  \bibfield  {author} {\bibinfo {author} {\bibfnamefont {J.}~\bibnamefont {Frauni{\'e}}}, \bibinfo {author} {\bibfnamefont {R.}~\bibnamefont {Jamil}}, \bibinfo {author} {\bibfnamefont {R.}~\bibnamefont {Kantelberg}}, \bibinfo {author} {\bibfnamefont {S.}~\bibnamefont {Roux}}, \bibinfo {author} {\bibfnamefont {L.}~\bibnamefont {Petit}}, \bibinfo {author} {\bibfnamefont {E.}~\bibnamefont {Lepleux}}, \bibinfo {author} {\bibfnamefont {L.}~\bibnamefont {Pacheco}}, \bibinfo {author} {\bibfnamefont {K.}~\bibnamefont {Watanabe}}, \bibinfo {author} {\bibfnamefont {T.}~\bibnamefont {Taniguchi}}, \bibinfo {author} {\bibfnamefont {V.}~\bibnamefont {Jacques}}, \bibinfo {author} {\bibfnamefont {L.}~\bibnamefont {Lombez}}, \bibinfo {author} {\bibfnamefont {M.~M.}\ \bibnamefont {Glazov}}, \bibinfo {author} {\bibfnamefont {B.}~\bibnamefont {Lassagne}}, \bibinfo {author} {\bibfnamefont {X.}~\bibnamefont {Marie}}, \ and\ \bibinfo {author} {\bibfnamefont {C.}~\bibnamefont {Robert}},\ }\href@noop {} {\bibfield  {journal}
  {\bibinfo  {journal} {Phys. Rev. Mater.}\ }\textbf {\bibinfo {volume} {7}},\ \bibinfo {pages} {L121002} (\bibinfo {year} {2023})}\BibitemShut {NoStop}%
\bibitem [{\citenamefont {Kim}\ \emph {et~al.}(2023)\citenamefont {Kim}, \citenamefont {Dominguez}, \citenamefont {Mayorga-Luna}, \citenamefont {Ye}, \citenamefont {Embley}, \citenamefont {Tan}, \citenamefont {Ni}, \citenamefont {Liu}, \citenamefont {Ford}, \citenamefont {Gao}, \citenamefont {Arash}, \citenamefont {Watanabe}, \citenamefont {Taniguchi}, \citenamefont {Kim}, \citenamefont {Shih}, \citenamefont {Lai}, \citenamefont {Yao}, \citenamefont {Yang}, \citenamefont {Li},\ and\ \citenamefont {Miyahara}}]{Kim2023-lg}%
  \BibitemOpen
  \bibfield  {author} {\bibinfo {author} {\bibfnamefont {D.~S.}\ \bibnamefont {Kim}}, \bibinfo {author} {\bibfnamefont {R.~C.}\ \bibnamefont {Dominguez}}, \bibinfo {author} {\bibfnamefont {R.}~\bibnamefont {Mayorga-Luna}}, \bibinfo {author} {\bibfnamefont {D.}~\bibnamefont {Ye}}, \bibinfo {author} {\bibfnamefont {J.}~\bibnamefont {Embley}}, \bibinfo {author} {\bibfnamefont {T.}~\bibnamefont {Tan}}, \bibinfo {author} {\bibfnamefont {Y.}~\bibnamefont {Ni}}, \bibinfo {author} {\bibfnamefont {Z.}~\bibnamefont {Liu}}, \bibinfo {author} {\bibfnamefont {M.}~\bibnamefont {Ford}}, \bibinfo {author} {\bibfnamefont {F.~Y.}\ \bibnamefont {Gao}}, \bibinfo {author} {\bibfnamefont {S.}~\bibnamefont {Arash}}, \bibinfo {author} {\bibfnamefont {K.}~\bibnamefont {Watanabe}}, \bibinfo {author} {\bibfnamefont {T.}~\bibnamefont {Taniguchi}}, \bibinfo {author} {\bibfnamefont {S.}~\bibnamefont {Kim}}, \bibinfo {author} {\bibfnamefont {C.-K.}\ \bibnamefont {Shih}}, \bibinfo {author} {\bibfnamefont {K.}~\bibnamefont {Lai}}, \bibinfo
  {author} {\bibfnamefont {W.}~\bibnamefont {Yao}}, \bibinfo {author} {\bibfnamefont {L.}~\bibnamefont {Yang}}, \bibinfo {author} {\bibfnamefont {X.}~\bibnamefont {Li}}, \ and\ \bibinfo {author} {\bibfnamefont {Y.}~\bibnamefont {Miyahara}},\ }\href@noop {} {\bibfield  {journal} {\bibinfo  {journal} {Nat. Mater.}\ } (\bibinfo {year} {2023})}\BibitemShut {NoStop}%
\bibitem [{\citenamefont {Yang}\ \emph {et~al.}(2020)\citenamefont {Yang}, \citenamefont {Li}, \citenamefont {Yin}, \citenamefont {Xu}, \citenamefont {Mullan}, \citenamefont {Taniguchi}, \citenamefont {Watanabe}, \citenamefont {Geim}, \citenamefont {Novoselov},\ and\ \citenamefont {Mishchenko}}]{Yang2020-gk}%
  \BibitemOpen
  \bibfield  {author} {\bibinfo {author} {\bibfnamefont {Y.}~\bibnamefont {Yang}}, \bibinfo {author} {\bibfnamefont {J.}~\bibnamefont {Li}}, \bibinfo {author} {\bibfnamefont {J.}~\bibnamefont {Yin}}, \bibinfo {author} {\bibfnamefont {S.}~\bibnamefont {Xu}}, \bibinfo {author} {\bibfnamefont {C.}~\bibnamefont {Mullan}}, \bibinfo {author} {\bibfnamefont {T.}~\bibnamefont {Taniguchi}}, \bibinfo {author} {\bibfnamefont {K.}~\bibnamefont {Watanabe}}, \bibinfo {author} {\bibfnamefont {A.~K.}\ \bibnamefont {Geim}}, \bibinfo {author} {\bibfnamefont {K.~S.}\ \bibnamefont {Novoselov}}, \ and\ \bibinfo {author} {\bibfnamefont {A.}~\bibnamefont {Mishchenko}},\ }\href@noop {} {\bibfield  {journal} {\bibinfo  {journal} {Sci Adv}\ }\textbf {\bibinfo {volume} {6}} (\bibinfo {year} {2020})}\BibitemShut {NoStop}%
\bibitem [{\citenamefont {Du}\ \emph {et~al.}(2017)\citenamefont {Du}, \citenamefont {Yu}, \citenamefont {Liao}, \citenamefont {Wang}, \citenamefont {Xie}, \citenamefont {Lu}, \citenamefont {Zhu}, \citenamefont {Li}, \citenamefont {Shen}, \citenamefont {Chen}, \citenamefont {Yang}, \citenamefont {Shi},\ and\ \citenamefont {Zhang}}]{Du2017-eg}%
  \BibitemOpen
  \bibfield  {author} {\bibinfo {author} {\bibfnamefont {L.}~\bibnamefont {Du}}, \bibinfo {author} {\bibfnamefont {H.}~\bibnamefont {Yu}}, \bibinfo {author} {\bibfnamefont {M.}~\bibnamefont {Liao}}, \bibinfo {author} {\bibfnamefont {S.}~\bibnamefont {Wang}}, \bibinfo {author} {\bibfnamefont {L.}~\bibnamefont {Xie}}, \bibinfo {author} {\bibfnamefont {X.}~\bibnamefont {Lu}}, \bibinfo {author} {\bibfnamefont {J.}~\bibnamefont {Zhu}}, \bibinfo {author} {\bibfnamefont {N.}~\bibnamefont {Li}}, \bibinfo {author} {\bibfnamefont {C.}~\bibnamefont {Shen}}, \bibinfo {author} {\bibfnamefont {P.}~\bibnamefont {Chen}}, \bibinfo {author} {\bibfnamefont {R.}~\bibnamefont {Yang}}, \bibinfo {author} {\bibfnamefont {D.}~\bibnamefont {Shi}}, \ and\ \bibinfo {author} {\bibfnamefont {G.}~\bibnamefont {Zhang}},\ }\href@noop {} {\bibfield  {journal} {\bibinfo  {journal} {Appl. Phys. Lett.}\ }\textbf {\bibinfo {volume} {111}},\ \bibinfo {pages} {263106} (\bibinfo {year} {2017})}\BibitemShut {NoStop}%
\bibitem [{\citenamefont {Wang}\ \emph {et~al.}(2016)\citenamefont {Wang}, \citenamefont {Chen}, \citenamefont {Li}, \citenamefont {Cheng}, \citenamefont {Yang}, \citenamefont {Wu}, \citenamefont {Xie}, \citenamefont {Zhang}, \citenamefont {Zhao}, \citenamefont {Lu}, \citenamefont {Chen}, \citenamefont {Wang}, \citenamefont {Meng}, \citenamefont {Tang}, \citenamefont {Yang}, \citenamefont {He}, \citenamefont {Liu}, \citenamefont {Shi}, \citenamefont {Watanabe}, \citenamefont {Taniguchi}, \citenamefont {Feng}, \citenamefont {Zhang},\ and\ \citenamefont {Zhang}}]{Wang2016-yj}%
  \BibitemOpen
  \bibfield  {author} {\bibinfo {author} {\bibfnamefont {D.}~\bibnamefont {Wang}}, \bibinfo {author} {\bibfnamefont {G.}~\bibnamefont {Chen}}, \bibinfo {author} {\bibfnamefont {C.}~\bibnamefont {Li}}, \bibinfo {author} {\bibfnamefont {M.}~\bibnamefont {Cheng}}, \bibinfo {author} {\bibfnamefont {W.}~\bibnamefont {Yang}}, \bibinfo {author} {\bibfnamefont {S.}~\bibnamefont {Wu}}, \bibinfo {author} {\bibfnamefont {G.}~\bibnamefont {Xie}}, \bibinfo {author} {\bibfnamefont {J.}~\bibnamefont {Zhang}}, \bibinfo {author} {\bibfnamefont {J.}~\bibnamefont {Zhao}}, \bibinfo {author} {\bibfnamefont {X.}~\bibnamefont {Lu}}, \bibinfo {author} {\bibfnamefont {P.}~\bibnamefont {Chen}}, \bibinfo {author} {\bibfnamefont {G.}~\bibnamefont {Wang}}, \bibinfo {author} {\bibfnamefont {J.}~\bibnamefont {Meng}}, \bibinfo {author} {\bibfnamefont {J.}~\bibnamefont {Tang}}, \bibinfo {author} {\bibfnamefont {R.}~\bibnamefont {Yang}}, \bibinfo {author} {\bibfnamefont {C.}~\bibnamefont {He}}, \bibinfo {author} {\bibfnamefont
  {D.}~\bibnamefont {Liu}}, \bibinfo {author} {\bibfnamefont {D.}~\bibnamefont {Shi}}, \bibinfo {author} {\bibfnamefont {K.}~\bibnamefont {Watanabe}}, \bibinfo {author} {\bibfnamefont {T.}~\bibnamefont {Taniguchi}}, \bibinfo {author} {\bibfnamefont {J.}~\bibnamefont {Feng}}, \bibinfo {author} {\bibfnamefont {Y.}~\bibnamefont {Zhang}}, \ and\ \bibinfo {author} {\bibfnamefont {G.}~\bibnamefont {Zhang}},\ }\href@noop {} {\bibfield  {journal} {\bibinfo  {journal} {Phys. Rev. Lett.}\ }\textbf {\bibinfo {volume} {116}},\ \bibinfo {pages} {126101} (\bibinfo {year} {2016})}\BibitemShut {NoStop}%
\bibitem [{\citenamefont {Ribeiro-Palau}\ \emph {et~al.}(2018)\citenamefont {Ribeiro-Palau}, \citenamefont {Zhang}, \citenamefont {Watanabe}, \citenamefont {Taniguchi}, \citenamefont {Hone},\ and\ \citenamefont {Dean}}]{Ribeiro-Palau2018-mu}%
  \BibitemOpen
  \bibfield  {author} {\bibinfo {author} {\bibfnamefont {R.}~\bibnamefont {Ribeiro-Palau}}, \bibinfo {author} {\bibfnamefont {C.}~\bibnamefont {Zhang}}, \bibinfo {author} {\bibfnamefont {K.}~\bibnamefont {Watanabe}}, \bibinfo {author} {\bibfnamefont {T.}~\bibnamefont {Taniguchi}}, \bibinfo {author} {\bibfnamefont {J.}~\bibnamefont {Hone}}, \ and\ \bibinfo {author} {\bibfnamefont {C.~R.}\ \bibnamefont {Dean}},\ }\href@noop {} {\bibfield  {journal} {\bibinfo  {journal} {Science}\ }\textbf {\bibinfo {volume} {361}},\ \bibinfo {pages} {690} (\bibinfo {year} {2018})}\BibitemShut {NoStop}%
\bibitem [{\citenamefont {Kapfer}\ \emph {et~al.}(2023)\citenamefont {Kapfer}, \citenamefont {Jessen}, \citenamefont {Eisele}, \citenamefont {Fu}, \citenamefont {Danielsen}, \citenamefont {Darlington}, \citenamefont {Moore}, \citenamefont {Finney}, \citenamefont {Marchese}, \citenamefont {Hsieh}, \citenamefont {Majchrzak}, \citenamefont {Jiang}, \citenamefont {Biswas}, \citenamefont {Dudin}, \citenamefont {Avila}, \citenamefont {Watanabe}, \citenamefont {Taniguchi}, \citenamefont {Ulstrup}, \citenamefont {B{\o}ggild}, \citenamefont {Schuck}, \citenamefont {Basov}, \citenamefont {Hone},\ and\ \citenamefont {Dean}}]{Kapfer2023-ji}%
  \BibitemOpen
  \bibfield  {author} {\bibinfo {author} {\bibfnamefont {M.}~\bibnamefont {Kapfer}}, \bibinfo {author} {\bibfnamefont {B.~S.}\ \bibnamefont {Jessen}}, \bibinfo {author} {\bibfnamefont {M.~E.}\ \bibnamefont {Eisele}}, \bibinfo {author} {\bibfnamefont {M.}~\bibnamefont {Fu}}, \bibinfo {author} {\bibfnamefont {D.~R.}\ \bibnamefont {Danielsen}}, \bibinfo {author} {\bibfnamefont {T.~P.}\ \bibnamefont {Darlington}}, \bibinfo {author} {\bibfnamefont {S.~L.}\ \bibnamefont {Moore}}, \bibinfo {author} {\bibfnamefont {N.~R.}\ \bibnamefont {Finney}}, \bibinfo {author} {\bibfnamefont {A.}~\bibnamefont {Marchese}}, \bibinfo {author} {\bibfnamefont {V.}~\bibnamefont {Hsieh}}, \bibinfo {author} {\bibfnamefont {P.}~\bibnamefont {Majchrzak}}, \bibinfo {author} {\bibfnamefont {Z.}~\bibnamefont {Jiang}}, \bibinfo {author} {\bibfnamefont {D.}~\bibnamefont {Biswas}}, \bibinfo {author} {\bibfnamefont {P.}~\bibnamefont {Dudin}}, \bibinfo {author} {\bibfnamefont {J.}~\bibnamefont {Avila}}, \bibinfo {author} {\bibfnamefont
  {K.}~\bibnamefont {Watanabe}}, \bibinfo {author} {\bibfnamefont {T.}~\bibnamefont {Taniguchi}}, \bibinfo {author} {\bibfnamefont {S.}~\bibnamefont {Ulstrup}}, \bibinfo {author} {\bibfnamefont {P.}~\bibnamefont {B{\o}ggild}}, \bibinfo {author} {\bibfnamefont {P.~J.}\ \bibnamefont {Schuck}}, \bibinfo {author} {\bibfnamefont {D.~N.}\ \bibnamefont {Basov}}, \bibinfo {author} {\bibfnamefont {J.}~\bibnamefont {Hone}}, \ and\ \bibinfo {author} {\bibfnamefont {C.~R.}\ \bibnamefont {Dean}},\ }\href@noop {} {\bibfield  {journal} {\bibinfo  {journal} {Science}\ }\textbf {\bibinfo {volume} {381}},\ \bibinfo {pages} {677} (\bibinfo {year} {2023})}\BibitemShut {NoStop}%
\bibitem [{\citenamefont {Souibgui}\ \emph {et~al.}(2017)\citenamefont {Souibgui}, \citenamefont {Ajlani}, \citenamefont {Cavanna}, \citenamefont {Oueslati}, \citenamefont {Meftah},\ and\ \citenamefont {Madouri}}]{Souibgui2017-ma}%
  \BibitemOpen
  \bibfield  {author} {\bibinfo {author} {\bibfnamefont {M.}~\bibnamefont {Souibgui}}, \bibinfo {author} {\bibfnamefont {H.}~\bibnamefont {Ajlani}}, \bibinfo {author} {\bibfnamefont {A.}~\bibnamefont {Cavanna}}, \bibinfo {author} {\bibfnamefont {M.}~\bibnamefont {Oueslati}}, \bibinfo {author} {\bibfnamefont {A.}~\bibnamefont {Meftah}}, \ and\ \bibinfo {author} {\bibfnamefont {A.}~\bibnamefont {Madouri}},\ }\href@noop {} {\bibfield  {journal} {\bibinfo  {journal} {Superlattices Microstruct.}\ }\textbf {\bibinfo {volume} {112}},\ \bibinfo {pages} {394} (\bibinfo {year} {2017})}\BibitemShut {NoStop}%
\bibitem [{\citenamefont {Baek}\ \emph {et~al.}(2023)\citenamefont {Baek}, \citenamefont {Kim}, \citenamefont {Lim}, \citenamefont {Hong}, \citenamefont {Chang}, \citenamefont {Ryu}, \citenamefont {Jung}, \citenamefont {Jang}, \citenamefont {Kim}, \citenamefont {Zhang}, \citenamefont {Watanabe}, \citenamefont {Taniguchi}, \citenamefont {Huang}, \citenamefont {Cheong}, \citenamefont {Kim},\ and\ \citenamefont {Lee}}]{Baek2023-ic}%
  \BibitemOpen
  \bibfield  {author} {\bibinfo {author} {\bibfnamefont {J.-H.}\ \bibnamefont {Baek}}, \bibinfo {author} {\bibfnamefont {H.~G.}\ \bibnamefont {Kim}}, \bibinfo {author} {\bibfnamefont {S.~Y.}\ \bibnamefont {Lim}}, \bibinfo {author} {\bibfnamefont {S.~C.}\ \bibnamefont {Hong}}, \bibinfo {author} {\bibfnamefont {Y.}~\bibnamefont {Chang}}, \bibinfo {author} {\bibfnamefont {H.}~\bibnamefont {Ryu}}, \bibinfo {author} {\bibfnamefont {Y.}~\bibnamefont {Jung}}, \bibinfo {author} {\bibfnamefont {H.}~\bibnamefont {Jang}}, \bibinfo {author} {\bibfnamefont {J.}~\bibnamefont {Kim}}, \bibinfo {author} {\bibfnamefont {Y.}~\bibnamefont {Zhang}}, \bibinfo {author} {\bibfnamefont {K.}~\bibnamefont {Watanabe}}, \bibinfo {author} {\bibfnamefont {T.}~\bibnamefont {Taniguchi}}, \bibinfo {author} {\bibfnamefont {P.~Y.}\ \bibnamefont {Huang}}, \bibinfo {author} {\bibfnamefont {H.}~\bibnamefont {Cheong}}, \bibinfo {author} {\bibfnamefont {M.}~\bibnamefont {Kim}}, \ and\ \bibinfo {author} {\bibfnamefont {G.-H.}\ \bibnamefont {Lee}},\
  }\href@noop {} {\bibfield  {journal} {\bibinfo  {journal} {Nat. Mater.}\ ,\ \bibinfo {pages} {1}} (\bibinfo {year} {2023})}\BibitemShut {NoStop}%
\bibitem [{\citenamefont {Zhu}\ \emph {et~al.}(2018)\citenamefont {Zhu}, \citenamefont {Zhan}, \citenamefont {Shih}, \citenamefont {Wan}, \citenamefont {Lu}, \citenamefont {Huang}, \citenamefont {Guo}, \citenamefont {Zhou},\ and\ \citenamefont {Cai}}]{Zhu2018-ya}%
  \BibitemOpen
  \bibfield  {author} {\bibinfo {author} {\bibfnamefont {Z.}~\bibnamefont {Zhu}}, \bibinfo {author} {\bibfnamefont {L.}~\bibnamefont {Zhan}}, \bibinfo {author} {\bibfnamefont {T.-M.}\ \bibnamefont {Shih}}, \bibinfo {author} {\bibfnamefont {W.}~\bibnamefont {Wan}}, \bibinfo {author} {\bibfnamefont {J.}~\bibnamefont {Lu}}, \bibinfo {author} {\bibfnamefont {J.}~\bibnamefont {Huang}}, \bibinfo {author} {\bibfnamefont {S.}~\bibnamefont {Guo}}, \bibinfo {author} {\bibfnamefont {Y.}~\bibnamefont {Zhou}}, \ and\ \bibinfo {author} {\bibfnamefont {W.}~\bibnamefont {Cai}},\ }\href@noop {} {\bibfield  {journal} {\bibinfo  {journal} {Small}\ }\textbf {\bibinfo {volume} {14}},\ \bibinfo {pages} {e1802498} (\bibinfo {year} {2018})}\BibitemShut {NoStop}%
\bibitem [{\citenamefont {Alden}\ \emph {et~al.}(2013)\citenamefont {Alden}, \citenamefont {Tsen}, \citenamefont {Huang}, \citenamefont {Hovden}, \citenamefont {Brown}, \citenamefont {Park}, \citenamefont {Muller},\ and\ \citenamefont {McEuen}}]{Alden2013-mi}%
  \BibitemOpen
  \bibfield  {author} {\bibinfo {author} {\bibfnamefont {J.~S.}\ \bibnamefont {Alden}}, \bibinfo {author} {\bibfnamefont {A.~W.}\ \bibnamefont {Tsen}}, \bibinfo {author} {\bibfnamefont {P.~Y.}\ \bibnamefont {Huang}}, \bibinfo {author} {\bibfnamefont {R.}~\bibnamefont {Hovden}}, \bibinfo {author} {\bibfnamefont {L.}~\bibnamefont {Brown}}, \bibinfo {author} {\bibfnamefont {J.}~\bibnamefont {Park}}, \bibinfo {author} {\bibfnamefont {D.~A.}\ \bibnamefont {Muller}}, \ and\ \bibinfo {author} {\bibfnamefont {P.~L.}\ \bibnamefont {McEuen}},\ }\href@noop {} {\bibfield  {journal} {\bibinfo  {journal} {Proc. Natl. Acad. Sci. U. S. A.}\ }\textbf {\bibinfo {volume} {110}},\ \bibinfo {pages} {11256} (\bibinfo {year} {2013})}\BibitemShut {NoStop}%
\bibitem [{\citenamefont {Woods}\ \emph {et~al.}(2016)\citenamefont {Woods}, \citenamefont {Withers}, \citenamefont {Zhu}, \citenamefont {Cao}, \citenamefont {Yu}, \citenamefont {Kozikov}, \citenamefont {Ben~Shalom}, \citenamefont {Morozov}, \citenamefont {van Wijk}, \citenamefont {Fasolino}, \citenamefont {Katsnelson}, \citenamefont {Watanabe}, \citenamefont {Taniguchi}, \citenamefont {Geim}, \citenamefont {Mishchenko},\ and\ \citenamefont {Novoselov}}]{Woods2016-lb}%
  \BibitemOpen
  \bibfield  {author} {\bibinfo {author} {\bibfnamefont {C.~R.}\ \bibnamefont {Woods}}, \bibinfo {author} {\bibfnamefont {F.}~\bibnamefont {Withers}}, \bibinfo {author} {\bibfnamefont {M.~J.}\ \bibnamefont {Zhu}}, \bibinfo {author} {\bibfnamefont {Y.}~\bibnamefont {Cao}}, \bibinfo {author} {\bibfnamefont {G.}~\bibnamefont {Yu}}, \bibinfo {author} {\bibfnamefont {A.}~\bibnamefont {Kozikov}}, \bibinfo {author} {\bibfnamefont {M.}~\bibnamefont {Ben~Shalom}}, \bibinfo {author} {\bibfnamefont {S.~V.}\ \bibnamefont {Morozov}}, \bibinfo {author} {\bibfnamefont {M.~M.}\ \bibnamefont {van Wijk}}, \bibinfo {author} {\bibfnamefont {A.}~\bibnamefont {Fasolino}}, \bibinfo {author} {\bibfnamefont {M.~I.}\ \bibnamefont {Katsnelson}}, \bibinfo {author} {\bibfnamefont {K.}~\bibnamefont {Watanabe}}, \bibinfo {author} {\bibfnamefont {T.}~\bibnamefont {Taniguchi}}, \bibinfo {author} {\bibfnamefont {A.~K.}\ \bibnamefont {Geim}}, \bibinfo {author} {\bibfnamefont {A.}~\bibnamefont {Mishchenko}}, \ and\ \bibinfo {author}
  {\bibfnamefont {K.~S.}\ \bibnamefont {Novoselov}},\ }\href@noop {} {\bibfield  {journal} {\bibinfo  {journal} {Nat. Commun.}\ }\textbf {\bibinfo {volume} {7}},\ \bibinfo {pages} {10800} (\bibinfo {year} {2016})}\BibitemShut {NoStop}%
\bibitem [{\citenamefont {Wang}\ \emph {et~al.}(2015)\citenamefont {Wang}, \citenamefont {Gao}, \citenamefont {Wen}, \citenamefont {Han}, \citenamefont {Taniguchi}, \citenamefont {Watanabe}, \citenamefont {Koshino}, \citenamefont {Hone},\ and\ \citenamefont {Dean}}]{Wang2015-ly}%
  \BibitemOpen
  \bibfield  {author} {\bibinfo {author} {\bibfnamefont {L.}~\bibnamefont {Wang}}, \bibinfo {author} {\bibfnamefont {Y.}~\bibnamefont {Gao}}, \bibinfo {author} {\bibfnamefont {B.}~\bibnamefont {Wen}}, \bibinfo {author} {\bibfnamefont {Z.}~\bibnamefont {Han}}, \bibinfo {author} {\bibfnamefont {T.}~\bibnamefont {Taniguchi}}, \bibinfo {author} {\bibfnamefont {K.}~\bibnamefont {Watanabe}}, \bibinfo {author} {\bibfnamefont {M.}~\bibnamefont {Koshino}}, \bibinfo {author} {\bibfnamefont {J.}~\bibnamefont {Hone}}, \ and\ \bibinfo {author} {\bibfnamefont {C.~R.}\ \bibnamefont {Dean}},\ }\href@noop {} {\bibfield  {journal} {\bibinfo  {journal} {Science}\ }\textbf {\bibinfo {volume} {350}},\ \bibinfo {pages} {1231} (\bibinfo {year} {2015})}\BibitemShut {NoStop}%
\bibitem [{\citenamefont {de~Jong}\ \emph {et~al.}(2022)\citenamefont {de~Jong}, \citenamefont {Benschop}, \citenamefont {Chen}, \citenamefont {Krasovskii}, \citenamefont {de~Dood}, \citenamefont {Tromp}, \citenamefont {Allan},\ and\ \citenamefont {van~der Molen}}]{De_Jong2022-fo}%
  \BibitemOpen
  \bibfield  {author} {\bibinfo {author} {\bibfnamefont {T.~A.}\ \bibnamefont {de~Jong}}, \bibinfo {author} {\bibfnamefont {T.}~\bibnamefont {Benschop}}, \bibinfo {author} {\bibfnamefont {X.}~\bibnamefont {Chen}}, \bibinfo {author} {\bibfnamefont {E.~E.}\ \bibnamefont {Krasovskii}}, \bibinfo {author} {\bibfnamefont {M.~J.~A.}\ \bibnamefont {de~Dood}}, \bibinfo {author} {\bibfnamefont {R.~M.}\ \bibnamefont {Tromp}}, \bibinfo {author} {\bibfnamefont {M.~P.}\ \bibnamefont {Allan}}, \ and\ \bibinfo {author} {\bibfnamefont {S.~J.}\ \bibnamefont {van~der Molen}},\ }\href@noop {} {\bibfield  {journal} {\bibinfo  {journal} {Nat. Commun.}\ }\textbf {\bibinfo {volume} {13}},\ \bibinfo {pages} {70} (\bibinfo {year} {2022})}\BibitemShut {NoStop}%
\bibitem [{\citenamefont {Zhu}\ \emph {et~al.}(2016)\citenamefont {Zhu}, \citenamefont {Ghazaryan}, \citenamefont {Son}, \citenamefont {Woods}, \citenamefont {Misra}, \citenamefont {He}, \citenamefont {Taniguchi}, \citenamefont {Watanabe}, \citenamefont {Novoselov}, \citenamefont {Cao},\ and\ \citenamefont {Mishchenko}}]{Zhu2016-mq}%
  \BibitemOpen
  \bibfield  {author} {\bibinfo {author} {\bibfnamefont {M.}~\bibnamefont {Zhu}}, \bibinfo {author} {\bibfnamefont {D.}~\bibnamefont {Ghazaryan}}, \bibinfo {author} {\bibfnamefont {S.-K.}\ \bibnamefont {Son}}, \bibinfo {author} {\bibfnamefont {C.~R.}\ \bibnamefont {Woods}}, \bibinfo {author} {\bibfnamefont {A.}~\bibnamefont {Misra}}, \bibinfo {author} {\bibfnamefont {L.}~\bibnamefont {He}}, \bibinfo {author} {\bibfnamefont {T.}~\bibnamefont {Taniguchi}}, \bibinfo {author} {\bibfnamefont {K.}~\bibnamefont {Watanabe}}, \bibinfo {author} {\bibfnamefont {K.~S.}\ \bibnamefont {Novoselov}}, \bibinfo {author} {\bibfnamefont {Y.}~\bibnamefont {Cao}}, \ and\ \bibinfo {author} {\bibfnamefont {A.}~\bibnamefont {Mishchenko}},\ }\href@noop {} {\bibfield  {journal} {\bibinfo  {journal} {2D Mater.}\ }\textbf {\bibinfo {volume} {4}},\ \bibinfo {pages} {011013} (\bibinfo {year} {2016})}\BibitemShut {NoStop}%
\bibitem [{\citenamefont {Cosma}\ \emph {et~al.}(2014)\citenamefont {Cosma}, \citenamefont {Wallbank}, \citenamefont {Cheianov},\ and\ \citenamefont {Fal'ko}}]{Cosma2014-hl}%
  \BibitemOpen
  \bibfield  {author} {\bibinfo {author} {\bibfnamefont {D.~A.}\ \bibnamefont {Cosma}}, \bibinfo {author} {\bibfnamefont {J.~R.}\ \bibnamefont {Wallbank}}, \bibinfo {author} {\bibfnamefont {V.}~\bibnamefont {Cheianov}}, \ and\ \bibinfo {author} {\bibfnamefont {V.~I.}\ \bibnamefont {Fal'ko}},\ }\href@noop {} {\bibfield  {journal} {\bibinfo  {journal} {Faraday Discuss.}\ }\textbf {\bibinfo {volume} {173}},\ \bibinfo {pages} {137} (\bibinfo {year} {2014})}\BibitemShut {NoStop}%
\bibitem [{\citenamefont {Bai}\ \emph {et~al.}(2020)\citenamefont {Bai}, \citenamefont {Zhou}, \citenamefont {Wang}, \citenamefont {Wu}, \citenamefont {McGilly}, \citenamefont {Halbertal}, \citenamefont {Lo}, \citenamefont {Liu}, \citenamefont {Ardelean}, \citenamefont {Rivera}, \citenamefont {Finney}, \citenamefont {Yang}, \citenamefont {Basov}, \citenamefont {Yao}, \citenamefont {Xu}, \citenamefont {Hone}, \citenamefont {Pasupathy},\ and\ \citenamefont {Zhu}}]{Bai2020-wr}%
  \BibitemOpen
  \bibfield  {author} {\bibinfo {author} {\bibfnamefont {Y.}~\bibnamefont {Bai}}, \bibinfo {author} {\bibfnamefont {L.}~\bibnamefont {Zhou}}, \bibinfo {author} {\bibfnamefont {J.}~\bibnamefont {Wang}}, \bibinfo {author} {\bibfnamefont {W.}~\bibnamefont {Wu}}, \bibinfo {author} {\bibfnamefont {L.~J.}\ \bibnamefont {McGilly}}, \bibinfo {author} {\bibfnamefont {D.}~\bibnamefont {Halbertal}}, \bibinfo {author} {\bibfnamefont {C.~F.~B.}\ \bibnamefont {Lo}}, \bibinfo {author} {\bibfnamefont {F.}~\bibnamefont {Liu}}, \bibinfo {author} {\bibfnamefont {J.}~\bibnamefont {Ardelean}}, \bibinfo {author} {\bibfnamefont {P.}~\bibnamefont {Rivera}}, \bibinfo {author} {\bibfnamefont {N.~R.}\ \bibnamefont {Finney}}, \bibinfo {author} {\bibfnamefont {X.-C.}\ \bibnamefont {Yang}}, \bibinfo {author} {\bibfnamefont {D.~N.}\ \bibnamefont {Basov}}, \bibinfo {author} {\bibfnamefont {W.}~\bibnamefont {Yao}}, \bibinfo {author} {\bibfnamefont {X.}~\bibnamefont {Xu}}, \bibinfo {author} {\bibfnamefont {J.}~\bibnamefont {Hone}}, \bibinfo
  {author} {\bibfnamefont {A.~N.}\ \bibnamefont {Pasupathy}}, \ and\ \bibinfo {author} {\bibfnamefont {X.-Y.}\ \bibnamefont {Zhu}},\ }\href@noop {} {\bibfield  {journal} {\bibinfo  {journal} {Nat. Mater.}\ }\textbf {\bibinfo {volume} {19}},\ \bibinfo {pages} {1068} (\bibinfo {year} {2020})}\BibitemShut {NoStop}%
\bibitem [{\citenamefont {Ma}\ \emph {et~al.}(2017)\citenamefont {Ma}, \citenamefont {Chen}, \citenamefont {Arnold},\ and\ \citenamefont {Chu}}]{Ma2017-tx}%
  \BibitemOpen
  \bibfield  {author} {\bibinfo {author} {\bibfnamefont {C.}~\bibnamefont {Ma}}, \bibinfo {author} {\bibfnamefont {Y.}~\bibnamefont {Chen}}, \bibinfo {author} {\bibfnamefont {W.}~\bibnamefont {Arnold}}, \ and\ \bibinfo {author} {\bibfnamefont {J.}~\bibnamefont {Chu}},\ }\href@noop {} {\bibfield  {journal} {\bibinfo  {journal} {J. Appl. Phys.}\ }\textbf {\bibinfo {volume} {121}} (\bibinfo {year} {2017})}\BibitemShut {NoStop}%
\bibitem [{\citenamefont {van Es}\ \emph {et~al.}(2018)\citenamefont {van Es}, \citenamefont {Mohtashami}, \citenamefont {Thijssen}, \citenamefont {Piras}, \citenamefont {van Neer},\ and\ \citenamefont {Sadeghian}}]{Van_Es2018-lc}%
  \BibitemOpen
  \bibfield  {author} {\bibinfo {author} {\bibfnamefont {M.~H.}\ \bibnamefont {van Es}}, \bibinfo {author} {\bibfnamefont {A.}~\bibnamefont {Mohtashami}}, \bibinfo {author} {\bibfnamefont {R.~M.~T.}\ \bibnamefont {Thijssen}}, \bibinfo {author} {\bibfnamefont {D.}~\bibnamefont {Piras}}, \bibinfo {author} {\bibfnamefont {P.~L. M.~J.}\ \bibnamefont {van Neer}}, \ and\ \bibinfo {author} {\bibfnamefont {H.}~\bibnamefont {Sadeghian}},\ }\href@noop {} {\bibfield  {journal} {\bibinfo  {journal} {Ultramicroscopy}\ }\textbf {\bibinfo {volume} {184}},\ \bibinfo {pages} {209} (\bibinfo {year} {2018})}\BibitemShut {NoStop}%
\bibitem [{\citenamefont {Woods}\ \emph {et~al.}(2014)\citenamefont {Woods}, \citenamefont {Britnell}, \citenamefont {Eckmann}, \citenamefont {Ma}, \citenamefont {Lu}, \citenamefont {Guo}, \citenamefont {Lin}, \citenamefont {Yu}, \citenamefont {Cao}, \citenamefont {Gorbachev}, \citenamefont {Kretinin}, \citenamefont {Park}, \citenamefont {Ponomarenko}, \citenamefont {Katsnelson}, \citenamefont {Gornostyrev}, \citenamefont {Watanabe}, \citenamefont {Taniguchi}, \citenamefont {Casiraghi}, \citenamefont {Gao}, \citenamefont {Geim},\ and\ \citenamefont {Novoselov}}]{Woods2014-bf}%
  \BibitemOpen
  \bibfield  {author} {\bibinfo {author} {\bibfnamefont {C.~R.}\ \bibnamefont {Woods}}, \bibinfo {author} {\bibfnamefont {L.}~\bibnamefont {Britnell}}, \bibinfo {author} {\bibfnamefont {A.}~\bibnamefont {Eckmann}}, \bibinfo {author} {\bibfnamefont {R.~S.}\ \bibnamefont {Ma}}, \bibinfo {author} {\bibfnamefont {J.~C.}\ \bibnamefont {Lu}}, \bibinfo {author} {\bibfnamefont {H.~M.}\ \bibnamefont {Guo}}, \bibinfo {author} {\bibfnamefont {X.}~\bibnamefont {Lin}}, \bibinfo {author} {\bibfnamefont {G.~L.}\ \bibnamefont {Yu}}, \bibinfo {author} {\bibfnamefont {Y.}~\bibnamefont {Cao}}, \bibinfo {author} {\bibfnamefont {R.~V.}\ \bibnamefont {Gorbachev}}, \bibinfo {author} {\bibfnamefont {A.~V.}\ \bibnamefont {Kretinin}}, \bibinfo {author} {\bibfnamefont {J.}~\bibnamefont {Park}}, \bibinfo {author} {\bibfnamefont {L.~A.}\ \bibnamefont {Ponomarenko}}, \bibinfo {author} {\bibfnamefont {M.~I.}\ \bibnamefont {Katsnelson}}, \bibinfo {author} {\bibfnamefont {Y.~N.}\ \bibnamefont {Gornostyrev}}, \bibinfo {author} {\bibfnamefont
  {K.}~\bibnamefont {Watanabe}}, \bibinfo {author} {\bibfnamefont {T.}~\bibnamefont {Taniguchi}}, \bibinfo {author} {\bibfnamefont {C.}~\bibnamefont {Casiraghi}}, \bibinfo {author} {\bibfnamefont {H.-J.}\ \bibnamefont {Gao}}, \bibinfo {author} {\bibfnamefont {A.~K.}\ \bibnamefont {Geim}}, \ and\ \bibinfo {author} {\bibfnamefont {K.~S.}\ \bibnamefont {Novoselov}},\ }\href@noop {} {\bibfield  {journal} {\bibinfo  {journal} {Nat. Phys.}\ }\textbf {\bibinfo {volume} {10}},\ \bibinfo {pages} {451} (\bibinfo {year} {2014})}\BibitemShut {NoStop}%
\bibitem [{\citenamefont {Tamayo}\ and\ \citenamefont {Garc{\'i}a}(1997)}]{Tamayo1997-to}%
  \BibitemOpen
  \bibfield  {author} {\bibinfo {author} {\bibfnamefont {J.}~\bibnamefont {Tamayo}}\ and\ \bibinfo {author} {\bibfnamefont {R.}~\bibnamefont {Garc{\'i}a}},\ }\href@noop {} {\bibfield  {journal} {\bibinfo  {journal} {Appl. Phys. Lett.}\ }\textbf {\bibinfo {volume} {71}},\ \bibinfo {pages} {2394} (\bibinfo {year} {1997})}\BibitemShut {NoStop}%
\bibitem [{\citenamefont {Wang}\ \emph {et~al.}(2023)\citenamefont {Wang}, \citenamefont {Finney}, \citenamefont {Sharpe}, \citenamefont {Rodenbach}, \citenamefont {Hsueh}, \citenamefont {Watanabe}, \citenamefont {Taniguchi}, \citenamefont {Kastner}, \citenamefont {Vafek},\ and\ \citenamefont {Goldhaber-Gordon}}]{Wang2023-ns}%
  \BibitemOpen
  \bibfield  {author} {\bibinfo {author} {\bibfnamefont {X.}~\bibnamefont {Wang}}, \bibinfo {author} {\bibfnamefont {J.}~\bibnamefont {Finney}}, \bibinfo {author} {\bibfnamefont {A.~L.}\ \bibnamefont {Sharpe}}, \bibinfo {author} {\bibfnamefont {L.~K.}\ \bibnamefont {Rodenbach}}, \bibinfo {author} {\bibfnamefont {C.~L.}\ \bibnamefont {Hsueh}}, \bibinfo {author} {\bibfnamefont {K.}~\bibnamefont {Watanabe}}, \bibinfo {author} {\bibfnamefont {T.}~\bibnamefont {Taniguchi}}, \bibinfo {author} {\bibfnamefont {M.~A.}\ \bibnamefont {Kastner}}, \bibinfo {author} {\bibfnamefont {O.}~\bibnamefont {Vafek}}, \ and\ \bibinfo {author} {\bibfnamefont {D.}~\bibnamefont {Goldhaber-Gordon}},\ }\href@noop {} {\bibfield  {journal} {\bibinfo  {journal} {Proc. Natl. Acad. Sci. U. S. A.}\ }\textbf {\bibinfo {volume} {120}},\ \bibinfo {pages} {e2307151120} (\bibinfo {year} {2023})}\BibitemShut {NoStop}%
\bibitem [{\citenamefont {Peng}, \citenamefont {Ji},\ and\ \citenamefont {De}(2012)}]{Peng2012-zs}%
  \BibitemOpen
  \bibfield  {author} {\bibinfo {author} {\bibfnamefont {Q.}~\bibnamefont {Peng}}, \bibinfo {author} {\bibfnamefont {W.}~\bibnamefont {Ji}}, \ and\ \bibinfo {author} {\bibfnamefont {S.}~\bibnamefont {De}},\ }\href@noop {} {\bibfield  {journal} {\bibinfo  {journal} {Comput. Mater. Sci.}\ }\textbf {\bibinfo {volume} {56}},\ \bibinfo {pages} {11} (\bibinfo {year} {2012})}\BibitemShut {NoStop}%
\bibitem [{\citenamefont {Kazmierczak}\ \emph {et~al.}(2021)\citenamefont {Kazmierczak}, \citenamefont {Van~Winkle}, \citenamefont {Ophus}, \citenamefont {Bustillo}, \citenamefont {Carr}, \citenamefont {Brown}, \citenamefont {Ciston}, \citenamefont {Taniguchi}, \citenamefont {Watanabe},\ and\ \citenamefont {Bediako}}]{Kazmierczak2021-lm}%
  \BibitemOpen
  \bibfield  {author} {\bibinfo {author} {\bibfnamefont {N.~P.}\ \bibnamefont {Kazmierczak}}, \bibinfo {author} {\bibfnamefont {M.}~\bibnamefont {Van~Winkle}}, \bibinfo {author} {\bibfnamefont {C.}~\bibnamefont {Ophus}}, \bibinfo {author} {\bibfnamefont {K.~C.}\ \bibnamefont {Bustillo}}, \bibinfo {author} {\bibfnamefont {S.}~\bibnamefont {Carr}}, \bibinfo {author} {\bibfnamefont {H.~G.}\ \bibnamefont {Brown}}, \bibinfo {author} {\bibfnamefont {J.}~\bibnamefont {Ciston}}, \bibinfo {author} {\bibfnamefont {T.}~\bibnamefont {Taniguchi}}, \bibinfo {author} {\bibfnamefont {K.}~\bibnamefont {Watanabe}}, \ and\ \bibinfo {author} {\bibfnamefont {D.~K.}\ \bibnamefont {Bediako}},\ }\href@noop {} {\bibfield  {journal} {\bibinfo  {journal} {Nat. Mater.}\ }\textbf {\bibinfo {volume} {20}},\ \bibinfo {pages} {956} (\bibinfo {year} {2021})}\BibitemShut {NoStop}%
\bibitem [{\citenamefont {Engelke}\ \emph {et~al.}(2023)\citenamefont {Engelke}, \citenamefont {Yoo}, \citenamefont {Carr}, \citenamefont {Xu}, \citenamefont {Cazeaux}, \citenamefont {Allen}, \citenamefont {Valdivia}, \citenamefont {Luskin}, \citenamefont {Kaxiras}, \citenamefont {Kim}, \citenamefont {Han},\ and\ \citenamefont {Kim}}]{Engelke2023-ez}%
  \BibitemOpen
  \bibfield  {author} {\bibinfo {author} {\bibfnamefont {R.}~\bibnamefont {Engelke}}, \bibinfo {author} {\bibfnamefont {H.}~\bibnamefont {Yoo}}, \bibinfo {author} {\bibfnamefont {S.}~\bibnamefont {Carr}}, \bibinfo {author} {\bibfnamefont {K.}~\bibnamefont {Xu}}, \bibinfo {author} {\bibfnamefont {P.}~\bibnamefont {Cazeaux}}, \bibinfo {author} {\bibfnamefont {R.}~\bibnamefont {Allen}}, \bibinfo {author} {\bibfnamefont {A.~M.}\ \bibnamefont {Valdivia}}, \bibinfo {author} {\bibfnamefont {M.}~\bibnamefont {Luskin}}, \bibinfo {author} {\bibfnamefont {E.}~\bibnamefont {Kaxiras}}, \bibinfo {author} {\bibfnamefont {M.}~\bibnamefont {Kim}}, \bibinfo {author} {\bibfnamefont {J.~H.}\ \bibnamefont {Han}}, \ and\ \bibinfo {author} {\bibfnamefont {P.}~\bibnamefont {Kim}},\ }\href@noop {} {\bibfield  {journal} {\bibinfo  {journal} {Phys. Rev. B Condens. Matter}\ }\textbf {\bibinfo {volume} {107}},\ \bibinfo {pages} {125413} (\bibinfo {year} {2023})}\BibitemShut {NoStop}%
\bibitem [{\citenamefont {Van~Winkle}\ \emph {et~al.}(2023)\citenamefont {Van~Winkle}, \citenamefont {Craig}, \citenamefont {Carr}, \citenamefont {Dandu}, \citenamefont {Bustillo}, \citenamefont {Ciston}, \citenamefont {Ophus}, \citenamefont {Taniguchi}, \citenamefont {Watanabe}, \citenamefont {Raja}, \citenamefont {Griffin},\ and\ \citenamefont {Bediako}}]{Van_Winkle2023-ug}%
  \BibitemOpen
  \bibfield  {author} {\bibinfo {author} {\bibfnamefont {M.}~\bibnamefont {Van~Winkle}}, \bibinfo {author} {\bibfnamefont {I.~M.}\ \bibnamefont {Craig}}, \bibinfo {author} {\bibfnamefont {S.}~\bibnamefont {Carr}}, \bibinfo {author} {\bibfnamefont {M.}~\bibnamefont {Dandu}}, \bibinfo {author} {\bibfnamefont {K.~C.}\ \bibnamefont {Bustillo}}, \bibinfo {author} {\bibfnamefont {J.}~\bibnamefont {Ciston}}, \bibinfo {author} {\bibfnamefont {C.}~\bibnamefont {Ophus}}, \bibinfo {author} {\bibfnamefont {T.}~\bibnamefont {Taniguchi}}, \bibinfo {author} {\bibfnamefont {K.}~\bibnamefont {Watanabe}}, \bibinfo {author} {\bibfnamefont {A.}~\bibnamefont {Raja}}, \bibinfo {author} {\bibfnamefont {S.~M.}\ \bibnamefont {Griffin}}, \ and\ \bibinfo {author} {\bibfnamefont {D.~K.}\ \bibnamefont {Bediako}},\ }\href@noop {} {\bibfield  {journal} {\bibinfo  {journal} {Nat. Commun.}\ }\textbf {\bibinfo {volume} {14}},\ \bibinfo {pages} {2989} (\bibinfo {year} {2023})}\BibitemShut {NoStop}%
\bibitem [{\citenamefont {Liao}\ \emph {et~al.}(2022)\citenamefont {Liao}, \citenamefont {Nicolini}, \citenamefont {Du}, \citenamefont {Yuan}, \citenamefont {Wang}, \citenamefont {Yu}, \citenamefont {Tang}, \citenamefont {Cheng}, \citenamefont {Watanabe}, \citenamefont {Taniguchi}, \citenamefont {Gu}, \citenamefont {Claerbout}, \citenamefont {Silva}, \citenamefont {Kramer}, \citenamefont {Polcar}, \citenamefont {Yang}, \citenamefont {Shi},\ and\ \citenamefont {Zhang}}]{Liao2022-zl}%
  \BibitemOpen
  \bibfield  {author} {\bibinfo {author} {\bibfnamefont {M.}~\bibnamefont {Liao}}, \bibinfo {author} {\bibfnamefont {P.}~\bibnamefont {Nicolini}}, \bibinfo {author} {\bibfnamefont {L.}~\bibnamefont {Du}}, \bibinfo {author} {\bibfnamefont {J.}~\bibnamefont {Yuan}}, \bibinfo {author} {\bibfnamefont {S.}~\bibnamefont {Wang}}, \bibinfo {author} {\bibfnamefont {H.}~\bibnamefont {Yu}}, \bibinfo {author} {\bibfnamefont {J.}~\bibnamefont {Tang}}, \bibinfo {author} {\bibfnamefont {P.}~\bibnamefont {Cheng}}, \bibinfo {author} {\bibfnamefont {K.}~\bibnamefont {Watanabe}}, \bibinfo {author} {\bibfnamefont {T.}~\bibnamefont {Taniguchi}}, \bibinfo {author} {\bibfnamefont {L.}~\bibnamefont {Gu}}, \bibinfo {author} {\bibfnamefont {V.~E.~P.}\ \bibnamefont {Claerbout}}, \bibinfo {author} {\bibfnamefont {A.}~\bibnamefont {Silva}}, \bibinfo {author} {\bibfnamefont {D.}~\bibnamefont {Kramer}}, \bibinfo {author} {\bibfnamefont {T.}~\bibnamefont {Polcar}}, \bibinfo {author} {\bibfnamefont {R.}~\bibnamefont {Yang}}, \bibinfo
  {author} {\bibfnamefont {D.}~\bibnamefont {Shi}}, \ and\ \bibinfo {author} {\bibfnamefont {G.}~\bibnamefont {Zhang}},\ }\href@noop {} {\bibfield  {journal} {\bibinfo  {journal} {Nat. Mater.}\ }\textbf {\bibinfo {volume} {21}},\ \bibinfo {pages} {47} (\bibinfo {year} {2022})}\BibitemShut {NoStop}%
\bibitem [{\citenamefont {Dong}(2014)}]{Dong2014-iu}%
  \BibitemOpen
  \bibfield  {author} {\bibinfo {author} {\bibfnamefont {Y.}~\bibnamefont {Dong}},\ }\href@noop {} {\bibfield  {journal} {\bibinfo  {journal} {J. Phys. D Appl. Phys.}\ }\textbf {\bibinfo {volume} {47}},\ \bibinfo {pages} {055305} (\bibinfo {year} {2014})}\BibitemShut {NoStop}%
\bibitem [{\citenamefont {Dienwiebel}\ \emph {et~al.}(2004)\citenamefont {Dienwiebel}, \citenamefont {Verhoeven}, \citenamefont {Pradeep}, \citenamefont {Frenken}, \citenamefont {Heimberg},\ and\ \citenamefont {Zandbergen}}]{Dienwiebel2004-da}%
  \BibitemOpen
  \bibfield  {author} {\bibinfo {author} {\bibfnamefont {M.}~\bibnamefont {Dienwiebel}}, \bibinfo {author} {\bibfnamefont {G.~S.}\ \bibnamefont {Verhoeven}}, \bibinfo {author} {\bibfnamefont {N.}~\bibnamefont {Pradeep}}, \bibinfo {author} {\bibfnamefont {J.~W.~M.}\ \bibnamefont {Frenken}}, \bibinfo {author} {\bibfnamefont {J.~A.}\ \bibnamefont {Heimberg}}, \ and\ \bibinfo {author} {\bibfnamefont {H.~W.}\ \bibnamefont {Zandbergen}},\ }\href@noop {} {\bibfield  {journal} {\bibinfo  {journal} {Phys. Rev. Lett.}\ }\textbf {\bibinfo {volume} {92}},\ \bibinfo {pages} {126101} (\bibinfo {year} {2004})}\BibitemShut {NoStop}%
\bibitem [{\citenamefont {Zhang}\ \emph {et~al.}(2019)\citenamefont {Zhang}, \citenamefont {Hou}, \citenamefont {Li}, \citenamefont {Liu}, \citenamefont {Zhang}, \citenamefont {Feng},\ and\ \citenamefont {Li}}]{Zhang2019-fb}%
  \BibitemOpen
  \bibfield  {author} {\bibinfo {author} {\bibfnamefont {S.}~\bibnamefont {Zhang}}, \bibinfo {author} {\bibfnamefont {Y.}~\bibnamefont {Hou}}, \bibinfo {author} {\bibfnamefont {S.}~\bibnamefont {Li}}, \bibinfo {author} {\bibfnamefont {L.}~\bibnamefont {Liu}}, \bibinfo {author} {\bibfnamefont {Z.}~\bibnamefont {Zhang}}, \bibinfo {author} {\bibfnamefont {X.-Q.}\ \bibnamefont {Feng}}, \ and\ \bibinfo {author} {\bibfnamefont {Q.}~\bibnamefont {Li}},\ }\href@noop {} {\bibfield  {journal} {\bibinfo  {journal} {Proc. Natl. Acad. Sci. U. S. A.}\ }\textbf {\bibinfo {volume} {116}},\ \bibinfo {pages} {24452} (\bibinfo {year} {2019})}\BibitemShut {NoStop}%
\bibitem [{\citenamefont {Martin}\ \emph {et~al.}(1993)\citenamefont {Martin}, \citenamefont {Donnet}, \citenamefont {{Le Mogne T}},\ and\ \citenamefont {Epicier}}]{Martin1993-ge}%
  \BibitemOpen
  \bibfield  {author} {\bibinfo {author} {\bibfnamefont {J.~M.}\ \bibnamefont {Martin}}, \bibinfo {author} {\bibfnamefont {C.}~\bibnamefont {Donnet}}, \bibinfo {author} {\bibnamefont {{Le Mogne T}}}, \ and\ \bibinfo {author} {\bibfnamefont {T.}~\bibnamefont {Epicier}},\ }\href@noop {} {\bibfield  {journal} {\bibinfo  {journal} {Phys. Rev. B Condens. Matter}\ }\textbf {\bibinfo {volume} {48}},\ \bibinfo {pages} {10583} (\bibinfo {year} {1993})}\BibitemShut {NoStop}%
\bibitem [{\citenamefont {He}, \citenamefont {Muser},\ and\ \citenamefont {Robbins}(1999)}]{He1999-ed}%
  \BibitemOpen
  \bibfield  {author} {\bibinfo {author} {\bibfnamefont {G.}~\bibnamefont {He}}, \bibinfo {author} {\bibfnamefont {M.~H.}\ \bibnamefont {Muser}}, \ and\ \bibinfo {author} {\bibfnamefont {M.~O.}\ \bibnamefont {Robbins}},\ }\href@noop {} {\bibfield  {journal} {\bibinfo  {journal} {Science}\ }\textbf {\bibinfo {volume} {284}},\ \bibinfo {pages} {1650} (\bibinfo {year} {1999})}\BibitemShut {NoStop}%
\bibitem [{\citenamefont {Zhu}, \citenamefont {Pochet},\ and\ \citenamefont {Johnson}(2019)}]{Zhu2019-ay}%
  \BibitemOpen
  \bibfield  {author} {\bibinfo {author} {\bibfnamefont {S.}~\bibnamefont {Zhu}}, \bibinfo {author} {\bibfnamefont {P.}~\bibnamefont {Pochet}}, \ and\ \bibinfo {author} {\bibfnamefont {H.~T.}\ \bibnamefont {Johnson}},\ }\href@noop {} {\bibfield  {journal} {\bibinfo  {journal} {ACS Nano}\ }\textbf {\bibinfo {volume} {13}},\ \bibinfo {pages} {6925} (\bibinfo {year} {2019})}\BibitemShut {NoStop}%
\bibitem [{\citenamefont {Bagchi}, \citenamefont {Johnson},\ and\ \citenamefont {Chew}(2020)}]{Bagchi2020-jm}%
  \BibitemOpen
  \bibfield  {author} {\bibinfo {author} {\bibfnamefont {S.}~\bibnamefont {Bagchi}}, \bibinfo {author} {\bibfnamefont {H.~T.}\ \bibnamefont {Johnson}}, \ and\ \bibinfo {author} {\bibfnamefont {H.~B.}\ \bibnamefont {Chew}},\ }\href@noop {} {\bibfield  {journal} {\bibinfo  {journal} {Phys. Rev. B Condens. Matter}\ }\textbf {\bibinfo {volume} {101}},\ \bibinfo {pages} {054109} (\bibinfo {year} {2020})}\BibitemShut {NoStop}%
\bibitem [{\citenamefont {Yang}\ and\ \citenamefont {Zhang}(2023)}]{Yang2023-qf}%
  \BibitemOpen
  \bibfield  {author} {\bibinfo {author} {\bibfnamefont {X.}~\bibnamefont {Yang}}\ and\ \bibinfo {author} {\bibfnamefont {B.}~\bibnamefont {Zhang}},\ }\href@noop {} {\bibfield  {journal} {\bibinfo  {journal} {Sci. Rep.}\ }\textbf {\bibinfo {volume} {13}},\ \bibinfo {pages} {4364} (\bibinfo {year} {2023})}\BibitemShut {NoStop}%
\bibitem [{\citenamefont {Lee}\ \emph {et~al.}(2017)\citenamefont {Lee}, \citenamefont {Ko}, \citenamefont {Choi}, \citenamefont {Hwang}, \citenamefont {Kim}, \citenamefont {Salmeron},\ and\ \citenamefont {Park}}]{Lee2017-xn}%
  \BibitemOpen
  \bibfield  {author} {\bibinfo {author} {\bibfnamefont {H.}~\bibnamefont {Lee}}, \bibinfo {author} {\bibfnamefont {J.-H.}\ \bibnamefont {Ko}}, \bibinfo {author} {\bibfnamefont {J.~S.}\ \bibnamefont {Choi}}, \bibinfo {author} {\bibfnamefont {J.~H.}\ \bibnamefont {Hwang}}, \bibinfo {author} {\bibfnamefont {Y.-H.}\ \bibnamefont {Kim}}, \bibinfo {author} {\bibfnamefont {M.}~\bibnamefont {Salmeron}}, \ and\ \bibinfo {author} {\bibfnamefont {J.~Y.}\ \bibnamefont {Park}},\ }\href@noop {} {\bibfield  {journal} {\bibinfo  {journal} {J. Phys. Chem. Lett.}\ }\textbf {\bibinfo {volume} {8}},\ \bibinfo {pages} {3482} (\bibinfo {year} {2017})}\BibitemShut {NoStop}%
\bibitem [{\citenamefont {Dollekamp}\ \emph {et~al.}(2019)\citenamefont {Dollekamp}, \citenamefont {Bampoulis}, \citenamefont {Siekman}, \citenamefont {Kooij},\ and\ \citenamefont {Zandvliet}}]{Dollekamp2019-fh}%
  \BibitemOpen
  \bibfield  {author} {\bibinfo {author} {\bibfnamefont {E.}~\bibnamefont {Dollekamp}}, \bibinfo {author} {\bibfnamefont {P.}~\bibnamefont {Bampoulis}}, \bibinfo {author} {\bibfnamefont {M.~H.}\ \bibnamefont {Siekman}}, \bibinfo {author} {\bibfnamefont {E.~S.}\ \bibnamefont {Kooij}}, \ and\ \bibinfo {author} {\bibfnamefont {H.~J.~W.}\ \bibnamefont {Zandvliet}},\ }\href@noop {} {\bibfield  {journal} {\bibinfo  {journal} {Langmuir}\ }\textbf {\bibinfo {volume} {35}},\ \bibinfo {pages} {4886} (\bibinfo {year} {2019})}\BibitemShut {NoStop}%
\end{thebibliography}%


%

\end{document}